\documentstyle[referee]{mn}

\newif\ifAMStwofonts

\newcommand{\alt}{\la}
\newcommand{\agt}{\ga}

\ifoldfss

  \newcommand{\itl}[1] {{\it #1}}
  \newcommand{\bld}[1] {{\bf #1}}
  \ifCUPmtlplainloaded \else
    \NewTextAlphabet{textbfit} {cmbxti10} {}
    \NewTextAlphabet{textbfss} {cmssbx10} {}
    \NewMathAlphabet{mathbfit} {cmbxti10} {} 
    \NewMathAlphabet{mathbfss} {cmssbx10} {} 
  \fi
  \ifAMStwofonts
    \ifCUPmtlplainloaded \else
      \NewSymbolFont{upmath} {eurm10}
      \NewSymbolFont{AMSa} {msam10}
      \NewMathSymbol{\upi}     {0}{upmath}{19}
      \NewMathSymbol{\umu}     {0}{upmath}{16}
      \NewMathSymbol{\upartial}{0}{upmath}{40}
      \NewMathSymbol{\leqslant}{3}{AMSa}{36}
      \NewMathSymbol{\geqslant}{3}{AMSa}{3E}

    \fi
  \fi
\fi 

\ifnfssone
  \newmathalphabet{\mathit}
  \addtoversion{normal}{\mathit}{cmr}{m}{it}
  \addtoversion{bold}{\mathit}{cmr}{bx}{it}

  \newcommand{\itl}[1] {\mathit{#1}}
  \newcommand{\bld}[1] {\mathbf{#1}}
  \newmathalphabet{\mathbfit} 
  \addtoversion{normal}{\mathbfit}{cmr}{bx}{it}
  \addtoversion{bold}{\mathbfit}{cmr}{bx}{it}
  \newmathalphabet{\mathbfss} 
  \addtoversion{normal}{\mathbfss}{cmss}{bx}{n}
  \addtoversion{bold}{\mathbfss}{cmss}{bx}{n}
  \ifAMStwofonts
    \ifCUPmtlplainloaded \else
      \UseAMStwoboldmath
      \makeatletter
      \new@mathgroup\upmath@group
      \define@mathgroup\mv@normal\upmath@group{eur}{m}{n}
      \define@mathgroup\mv@bold\upmath@group{eur}{b}{n}
      \edef\UPM{\hexnumber\upmath@group}
      \new@mathgroup\amsa@group
      \define@mathgroup\mv@normal\amsa@group{msa}{m}{n}
      \define@mathgroup\mv@bold\amsa@group{msa}{m}{n}
      \edef\AMSa{\hexnumber\amsa@group}
      \makeatother
      \mathchardef\upi="0\UPM19
      \mathchardef\umu="0\UPM16
      \mathchardef\upartial="0\UPM40
      \mathchardef\leqslant="3\AMSa36
      \mathchardef\geqslant="3\AMSa3E
    \fi
  \fi
\fi 

\ifnfsstwo

  \newcommand{\itl}[1] {\mathit{#1}}
  \newcommand{\bld}[1] {\mathbf{#1}}
  \DeclareMathAlphabet{\mathbfit}{OT1}{cmr}{bx}{it}
  \SetMathAlphabet\mathbfit{bold}{OT1}{cmr}{bx}{it}
  \DeclareMathAlphabet{\mathbfss}{OT1}{cmss}{bx}{n}
  \SetMathAlphabet\mathbfss{bold}{OT1}{cmss}{bx}{n}
  \ifAMStwofonts
    \ifCUPmtlplainloaded \else
      \DeclareSymbolFont{UPM}{U}{eur}{m}{n}
      \SetSymbolFont{UPM}{bold}{U}{eur}{b}{n}
      \DeclareSymbolFont{AMSa}{U}{msa}{m}{n}
      \DeclareMathSymbol{\upi}{0}{UPM}{"19}
      \DeclareMathSymbol{\umu}{0}{UPM}{"16}
      \DeclareMathSymbol{\upartial}{0}{UPM}{"40}
      \DeclareMathSymbol{\leqslant}{3}{AMSa}{"36}
      \DeclareMathSymbol{\geqslant}{3}{AMSa}{"3E}
    \fi
  \fi
\fi 

\ifCUPmtlplainloaded \else
  \ifAMStwofonts \else 
    \def\upi{\pi}
    \def\umu{\mu}
    \def\upartial{\partial}
  \fi
\fi

\title{Superbubbles in magnetized interstellar media\\
--- Blowout or confinement ? ---}
\author[K. Tomisaka]
       {Kohji Tomisaka\thanks{E-mail: tomisaka@ed.niigata-u.ac.jp.} \\
        Faculty of Education, Niigata University, 8050 Ikarashi-2, 
        Niigata 950-21, Japan}
\date{Accepted \hrulefill.
      Received \hrulefill;
      in original form \hrulefill}

\pagerange{\pageref{firstpage}--\pageref{lastpage}}
\pubyear{1997}

\begin{document}

\maketitle

\label{firstpage}

\begin{abstract}
The importance of the interstellar magnetic field is studied
 in relation to the evolutions of superbubbles
 with a three-dimensional (3D) numerical
 magnetohydrodynamical (MHD) simulation.
A superbubble is a large supernova remnant driven by sequential supernova
 explosions in an OB association.
Its evolution is affected by the density stratification in the galactic disk.
After the size reaches 2--3 times the density scale-height,
 the superbubble expands preferentially in the $z$-direction.
Finally it can punch out the gas disk (blow-out).
On the other hand, the magnetic field running parallel to the galactic disk
 has an effect to prevent from expanding in the direction perpendicular to the 
 field.
The density stratification and the magnetic fields have completely
 opposite effects on the evolution of the superbubble.
We present results of 3D MHD simulation in which both effects are included.
As a result, it is concluded that
when the magnetic field has a much larger scale-height
 than the density,
 even for a model that the bubble would blow out from the disk
 if the magnetic field were absent,
 the magnetic field with the strength
 of 5 $\mu$G can confine the bubble in $|z| \alt 300$ pc for $\simeq$20 Myr
 (confinement).
In a model that the field strength decreases in the halo
 in proportion to $B \propto \rho^{1/2}$, the superbubble eventually blows out
 like a model of $B=0$ even if the magnetic field in
 the mid-plane is as strong as $B=5\mu$G.
\end{abstract}


\section{Introduction}

A superbubble is a complex consisting of an OB association, surrounding 
 X-ray emitting hot gas, and a corresponding HI hole/shell.
Three examples are well-known in our Galaxy:
 Cygnus (Cash et al. 1980), Orion-Eridanus (Cowie,  Songaila \& York 1979;
 Reynolds \& Ogden 1979), and Gum nebula (Reynolds 1976).
Superbubbles are found as HI shells and holes in external galaxies, such as
 LMC (Meaburn 1980; Dopita, Mathewson \& Ford 1985),
 M31 (Brinks \& Bajaja 1986),
 M33 (Deul \& den Hartog 1990), M101 (Kamphuis, Sancisi \&
 van der Hulst 1991), and so on.
Since the sizes of these objects are in the range of 100 pc -- 1kpc,
 this can not be explained by a single supernova explosion
 [the size of an ordinary supernova remnant (SNR) is $\alt 50$pc].
The amount of energy required for such a superbubble reaches
 $5\times 10^{51} {\rm erg} - 10^{54} {\rm erg}$
 (Tenorio-Tagle \& Bodenheimer 1988).
There are two models proposed for formation of superbubbles: 
(1) a large SNR driven by sequential
 supernova explosions in an OB association and (2) a complex formed
 by a collision of a high-velocity cloud and the galactic disk
 (Tenorio-Tagle 1991).
Here, we confine ourselves to the first model and discuss the evolutions.
For review papers of this field, see Tenorio-Tagle \& Bohdenheimer (1988),
 Spitzer (1990), Tomisaka (1991), Bisnovatyi-Kogan \& Silich (1995). 

Here, we summarize the evolution very briefly.
After the size of bubble exceeds the density scale-height,
 the bubble becomes elongated in the direction perpendicular to the galactic
 disk (Tomisaka \& Ikeuchi 1986; Tenorio-Tagle, Bodenheimer 
 \& R\'{o}\.{z}yczka, 1987; MacLow \& McCray 1987).
It is shown that when the mechanical luminosity released by sequential
 supernova explosions is high, the expansion of the bubble to the halo
 is accelerated and the hot gas contained in it flows into the
 galactic halo (galactic fountain), finally.
However, the magnetic fields running parallel to the galactic disk
 prevents the gas from flowing in the vertical direction (perpendicular to
 the fields).

Tomisaka (1990) studied the adiabatic evolution of a superbubble
 in uniform magnetic fields and showed that the bubble driven by  
 a mechanical luminosity of $L_{\rm SN}\simeq 3\times 10^{37}
{\rm erg\,s^{-1}}$ is confined in the galactic disk by the effect of 
 the magnetic field provided its strength is as large as $B_0=5\mu$G.
This shows the superbubble is confined in the galactic disk,
 if (1) the magnetic fields have a large scale-height, $H_B$, and
 (2)  $L_{\rm SN}\simeq 3\times 10^{37}{\rm erg s^{-1}}$ and  $B_0\agt 5\mu$G.
However, this result may be affected by the assumption of the adiabatic gas. 
The interstellar magnetic fields seems to play an important role 
 in the evolution of a superbubble (see also Mineshige, Shibata,
 \& Shapiro 1993, Ferriere, MacLow, \& Zweibel 1991).
 
In the present paper, a full magnetohydrodynamical calculation
 has been done including the radiative cooling.
Further, the effect of  distributions of magnetic field strength
 is studied.
Plan of the present paper is as follows:
 in section 2 is given an analytical estimate to a threshold  mechanical
 luminosity under which the superbubble is confined in the galactic disk.
From this estimation, we choose the values for the parameters $L_{\rm SN}$,
 $B_0$, and $H_B$.
Numerical method and other assumptions are also presented in $\S$2.
Section 3 is for the numerical result,
 in which the effect of the magnetic fields is shown.
In section 4,  we will discuss the observability of the superbubble
 in external galaxies.
Using the evolution obtained, we will show that a large fraction of the
 interstellar space is occupied with superbubbles.
    
\section{Analytical models}

As a first step, the expansion law of a superbubble is studied with
 a simple approximation of spherical symmetry.
[Historically, it was done by Bruhweiler et al. (1980) and
 Tomisaka, Habe \& Ikeuchi (1981).]
A self-similar solution for a wind-blown bubble by Weaver et al.(1977)
 shows us the expansion law for the shock wave driven by a steady energy
 release as
\begin{eqnarray}
 R_s & = & 271{\rm pc}({ L_{\rm SN} }/
                           { 3\times 10^{37}{\rm erg~s}^{-1} })^{1/5} 
\nonumber \\ 
 & \times &  ({ \rho_0 }/{ 6\times 10^{-25} {\rm g~cm}^{-3} })^{-1/5}
            ({ t }/{ 10{\rm Myr} })^{3/5},
\end{eqnarray}
where $L_{\rm SN}$ is a mean mechanical luminosity ejected from supernovae,
 i.e., $L_{\rm SN} = E_0/\Delta \tau$ using the total energy ejected from
 an SN ($E_0$) and the mean interval of two SNe ($\Delta \tau$).
 
Since a typical size of the superbubble ($R_s$) is much larger
 than a density scale-height of the galactic disk, 
 $H\equiv \int^\infty_0 \rho(z)dz/\rho(0)$,
 the evolution should be affected by the density stratification
 in the $z-$direction.
This problem is now two-dimensional.
A frontier of this problem was Chevalier \& Gardner (1974),
 who studied the evolution of an SNR in the galactic halo
 using a 2D hydro-code,
 although a small number of zones could be used at that time as 20 $\times$ 39.
This kind of problem became attacked seriously after mid-1980's
 using large 2D hydro-codes (Tomisaka \& Ikeuchi 1986, 1988; 
 Tenorio-Tagle, Bohdenheimer \& R\'{o}\.{z}yczka 1987,
 Tenorio-Tagle,  R\'{o}\.{z}yczka \& Bohdenheimer 1990; 
 MacLow, McCray \& Norman  1989;
 Igumentshchev, Shustov \& Tutukov, 1990) 
 and semi-analytic approximations
 (MacLow \& McCray 1987; Bisnovatyi-Kogan, Blinnikov \& Silich 1989;
 Koo \& McKee 1990; 1992).

Summarizing results of the numerical studies above,
 it is concluded that in the first stage,
 the expansion is spherical as long as the 
 size is much smaller than the density scale-height.
However, when the $L_{\rm SN}$ is sufficiently large,
 a bubble enters a new phase, in which
 an expansion in the direction of the density gradient becomes
 accelerated.
In an exponential atmosphere as $\rho=\rho_0\exp(-|z|/H)$,
 Mac Low et al. (1989) 
 showed that the shock front is accelerated after it passes 
 $z\simeq 2.9H$.
Although the exact numerical factor is dependent on how the density 
 distributes, $\rho(z)$, the numerical simulations insure that
 the shock is accelerated upwardly, after
 the size is larger than $(2-3)\times H$.
Finally, the superbubble punches out the gas disk and a hot gas contained
 in the bubble flows into the halo of the galaxy.
 
How about a less-energetic bubble?
If the thermal pressure $p_0$ of the ambient interstellar medium (ISM)
 is dominant over the post-shock ram pressure, i.e.,
 $\rho_0v_s^2 \alt p_0$, before the break-out,
 the bubble seems to be confined in the disk.
Using equation (1), 
 the ram pressure becomes equal to the thermal  pressure at the age of
\begin{equation}
 t_P \simeq 79 {\rm Myr}L_{38}^{1/2}n_0^{3/4}p_{-12}^{-5/4},
\end{equation}
 where
 $L_{38}=L_{\rm SN}/10^{38}{\rm erg~s}^{-1}$,
 $p_{-12}=p_0/10^{-12} {\rm dyn~cm}^{-2}$,
 and $n_0$ represents the number density of ISM.
We assume here that the condition for the break-out is  $R_s(t_P) > \alpha H$
 (Koo \& McKee 1992).
Equating $R_s(t_P) = \alpha H$, we obtain a threshold luminosity above
 which the break-out occurs as:
\begin{equation}
 L_{\rm crit}= 0.59 \times 10^{37}{\rm erg~s}^{-1}
             (\alpha^2/5)H_2^2n_0^{-1/2}p_{-12}^{3/2}, 
\label{eqn:Lcr0}
\end{equation}
with $H_2=H/100{\rm pc}$. 
The value of $\alpha\simeq \sqrt{5}$ is taken from an estimate 
 by numerical results by Mac~Low \& McCray (1988).
Thus, if $L_{\rm SN}$ is much smaller than $L_{\rm crit}$,
 the bubble is eventually confined in the galactic gas disk.

\subsection{Effect of the interstellar magnetic fields}

The magnetic pressure in ISM is expected as
\begin{equation}
 p_{\rm mag}=B_0^2/8\pi \simeq 10^{-12}{\rm dyn~cm}^{-2}(B_0/5\mu {\rm G})^2.
\end{equation}
Taking both magnetic ($p_{\rm mag}\sim 10^{-12}{\rm dyn~cm}^{-2}$)
 and thermal pressures ($p_{\rm th}=nkT\sim 2400 k\sim 7\times 10^{-13}
{\rm dyn~cm}^{-2}$) into account in equation (\ref{eqn:Lcr0}),
 the critical luminosity becomes
\begin{eqnarray}
 L_{\rm crit}&\simeq & 3\times 10^{37}{\rm erg~s}^{-1}
                        (H/180 {\rm pc})^2   
                        (n_0/0.3{\rm cm}^{-3})^{-1/2}\nonumber \\ 
              &\times & (p_0/1.7\times 10^{-12}{\rm dyn~cm}^{-2})^{3/2},
\label{eqn:Lcr1}
\end{eqnarray}
where we take a typical value for $H$ as 180 pc according to a model
 by Dickey \& Lockman (1990).
This shows that the magnetic field seems to play a role 
 for a superbubble with $L_{\rm SN}\simeq 3\times 10^{37}{\rm erg~s}^{-1}$.
This corresponds to a mean interval between subsequent supernova
 explosions in an OB association of $\Delta \tau \sim 10^6$yr and
 this is not inconsistent with that estimated from SN rate in our galaxy
(Tomisaka et al. 1981).
That is,
 the superbubble might be 
 magnetically pressure-confined in the disk.

Another possibility for a superbubble to be confined in the disk is
 that the mechanical luminosity decreases before the shock front is 
 accelerated in the $z-$ direction.
Since the age when the size $R_s$ exceeds $\alpha H$ is equal to
 $\tau_{acc} \simeq 6{\rm Myr} (\alpha/\sqrt{5})^{5/3} H_2^{5/3}
  (L_{\rm SN}/ 3\times 10^{37}{\rm erg~s}^{-1})^{-1/3} 
  (\rho/6\times 10^{-25} {\rm g~cm}^{-3})^{1/3}$,
in the case that
 the lifetime of the active phase of an OB association $\tau_{\rm life}$
 is shorter than $\tau_{\rm acc}$,
 the expansion is stalled due to the decrease of the mechanical luminosity.
If we assume the sequential supernova explosions continue at least
 $\tau_{\rm life}\agt 40$Myr (e.g. Shull \& Saken 1995), 
 only for an extremely poor OB association as 
 $L_{\rm SN} \alt 1 \times 10^{35}{\rm erg~s^{-1}}
 (\tau_{\rm life}/40{\rm Myr})^{-3}
 (\alpha/\sqrt{5})^5 H_2^5 (\rho/6\times 10^{-25} {\rm g~cm}^{-3})$
 the superbubble burns out before the shock is accelerated 
 upwardly in the $z-$direction.

\subsection{Models and basic equations} 
 We choose parameters for interstellar medium as follows:
 assuming the disk-plane ($z=0$) number density $n_0=0.3{\rm cm}^{-3}$,
 the gas temperature $T_0=8000$K,
 these give a pressure on the disk-plane as 
 $p_0\simeq 7\times 10^{-13}{\rm dyn~cm}^{-2}$.  
The density distribution perpendicular to the disk is taken based on a model
 proposed by Dickey \& Lockman (1990) as
\begin{eqnarray}
  n(z)=&\frac{n_0}{0.566} \left[   0.395 \exp \left[ -\frac{1}{2}\left(\frac{z}{90 {\rm pc}}\right)^2\right]\right.\nonumber \\
      & + 0.107 \exp \left[ -\frac{1}{2}\left(\frac{z}{225 {\rm pc}}\right)^2\right]\nonumber \\
     & \left. + 0.064 \exp \left[\left. -\frac{|z|}{403 {\rm pc}} \right.\right] \right],
\end{eqnarray}
 this yields an effective scale-height as
 $H_{\rm eff}\equiv \int_0^\infty \rho dz / \rho_0\simeq 180{\rm pc}$.
Applying equation (\ref{eqn:Lcr1}),
 model parameters chosen here give a critical luminosity of
 $ L_{\rm crit}\sim 3\times 10^{37}{\rm erg~s}^{-1}$.
Parameters taken are summarized in table 1.
\begin{table*}
\centering
 \begin{minipage}{140mm}
\caption{Model Parameters}
\begin{tabular}{cccccccc}
\hline
Model \ \ \ & $n_0$ &  $B_0$ & $T_0$ & $L_{\rm SN}$  & $B_x(z)$ & 
 $t_{\rm final}$\footnote{Numerical run continues up to this age.} &
 $R_{\rm equiv}$\footnote{Equivalent radius, eq.(\ref{Requiv}),
 at $t_{\rm final}$.}  \\
  & (${\rm cm}^{-3}$) & ($\mu$G) & (K) &     (${\rm erg~s}^{-1})$& &
 (Myr) & (pc)\\
\hline
\hline
A \dotfill\ & 0.3 & 5  & 8000 & $3\times 10^{37}$ & const & 22 & {255}  \\
B \dotfill\ & 0.3 & 5  & 8000 & $3\times 10^{38}$ & const & 20 & {567}  \\
C \dotfill\ & 0.3 & 5  & 8000 & $3\times 10^{37}$ & $\propto \rho^{1/2}$ & 36 & {592} \\
D \dotfill\ & 0.3 & 3  & 8000 & $3\times 10^{37}$ & $\propto \rho^{1/2}$ & 30 & {563} \\
E \dotfill\ & 0.3 & 5  & 8000 & $1\times 10^{37}$ & $\propto \rho^{1/2}$ & 40 & {196} \\
F \dotfill\ & 0.3 & 5  & 8000 & $3\times 10^{38}$ & $\propto \rho^{1/2}$ & 11 & {470} \\
Q \dotfill\ & 0.3 & 0  & 8000 & $3\times 10^{37}$ & no B & 50 & {1150} \\
\hline
\end{tabular}
\end{minipage}
\end{table*}
As an initial condition, we began with a hydrostatic state
 such as the thermal and magnetic pressures balance with the gravity due to
 the galactic disk as
\begin{equation}
  g_z(z)=\frac{1}{\rho}\frac{\partial (p_{\rm th}+B^2/8\pi)}{\partial z}.
\end{equation}
Actually,
 we assumed the distributions of gas and magnetic pressures first. 
Then, we chose the gravity consistently. 
In addition, we add a halo with constant density for $z > 0.8$ kpc
 to avoid numerical difficulties. 
The gravitational acceleration assumed here is somewhat different from that
is derived from observations of distributions and motions of 
K giants (Spitzer 1978).
For example, equation (7) gives $7\times 10^{-9} {\rm cm/s^2}$ for $z=200$ pc
 and $4\times 10^{-9} {\rm cm/s^2}$ for $z=400$ pc, while the 
 values derived from K giants for respective heights
 are equal to  $4\times 10^{-9} {\rm cm/s^2}$
 and $6\times 10^{-9} {\rm cm/s^2}$. 
Since the gravitational free-fall time scale $\sim (z/g_z)^{1/2}\sim
 20 {\rm Myr}(z/1 {\rm kpc})^{1/2}(g_z/8\times 10^{-9} {\rm cm/s^2})^{-1/2}$
 is comparable to a typical age of the superbubble,
 the gravitational acceleration may play a role in the late phase and then  
 the gravitational acceleration assumed here may underestimate the effect of
 gravitational  confinement of the superbubble.

The magnetic scale-height of our Galaxy can be estimated
 with the rotation measures of extragalactic radio sources (Spitzer 1978). 
This gives  a scale-height of a thicker component of the 
 magnetic disk whose e-folding scale-height is estimated as   
 $1.2\pm 0.4$ kpc (Han \& Qiao 1994 and references therein).
First, in Models A and B  we assume the strength of magnetic fields is constant
 ($B(z)=$const) irrespective of the height from the disk midplane, 
 in which  we will see the  effect of this thick magnetic disk.
In addition, from the rotation measure observations of galactic pulsars,
 it is shown there is an additional narrower component of the magnetic
 disk whose  Gaussian scale-height is as large as 170 pc (Thompson \& Nelson 1980).
To take this narrower component into consideration, we assume 
 the magnetic field strength decreases in 
 proportional to the square root of the gas density 
 in Models C--F.

Basic equations to be solved are the ideal MHD equations as follows:
\begin{equation}
	\frac{\partial \rho}{\partial t}+\nabla \cdot \rho \bld{v}=0,
		\label{eqn:cont}
\end{equation}
\begin{equation}
	\frac{\partial \rho\bld{v}}{\partial t}
		+\nabla \cdot \rho \bld{v}\bld{v}=
		-\nabla \itl{p} + \rho \bld{g} 
		+ \frac{1}{4\pi}\nabla\times\bld{B}\times\bld{B},
			\label{eqn:momentum}
\end{equation}
\begin{equation}
	\frac{\partial \epsilon}{\partial t}+\nabla \cdot \epsilon \bld{v}=
		 -\Lambda -p\nabla \bld{v},
			\label{eqn:thermal}
\end{equation}
with
\begin{equation}
	\epsilon=\frac{p}{\gamma-1},
\end{equation}
\begin{equation}
	\frac{\partial \bld{B}}{\partial t}=\nabla\times(\bld{v}\times\bld{B}),
		\label{eqn:induction}
\end{equation}
where $\Lambda(T,\rho)=L(T)n^2$ represents the radiative cooling.
Here, we fitted values derived by Raymond, Cox \& Smith(1976)
 with a third-order spline function.
Other symbols have usual meanings.
Equations (\ref{eqn:cont}), (\ref{eqn:momentum}), (\ref{eqn:thermal})
 and (\ref{eqn:induction}) represent, respectively,
 the continuity equation, the equation of motion,
 the energy conservation equation, and the induction equation under 
 the ideal MHD condition (with infinite conductivity). 
The adiabatic exponent $\gamma$ is taken equal to 5/3.

Numerical method we employ is 
 a finite-difference three-dimensional magnetohydrodynamics code.
Specifically, 
 a ``monotonic scheme'' (van Leer 1977; Norman \& Winkler 1986) is adopted
 to solve the hydrodynamics equation, equations (\ref{eqn:cont}), 
 (\ref{eqn:momentum}) and (\ref{eqn:thermal}).
On the other hand, a ``constrained transport'' method (Evans \& Hawley 1988) 
 and a ``method of characteristics'' (Stone \& Norman 1992)
 are used to solve the magnetic part of equation (\ref{eqn:momentum}) and
 the induction equation (\ref{eqn:induction}).
The code is well vectorized and parallelized.
Thus, it is executed very efficiently 
 on a vector-parallel machine Fujitsu VPP300 with 16 processors, e.g.,
 it needs only 0.9 sec -- 3.8 sec to proceed one time step 
 for grids 161$\times$161$\times$261
 -- 271$\times$271$\times$451, respectively.
The number of zones is increased just before the superbubble expands
 and it touches one of the numerical boundaries.
The grid spacing ($\Delta x$, $\Delta y$, $\Delta z$) is chosen 
 spatially constant as 5pc.
A superbubble is located at the origin $x=y=z=0$
 and energy and mass are ejected in a number of cells near the origin
 every time step $\Delta t$ to the amount of $L_{\rm SN}\Delta t$ and 
 $\dot{M}_{\rm SN}\Delta t$.
The mass ejection rate $\dot{M}_{\rm SN}$ is chosen as
 $3\times 10^{22}{\rm g~s^{-1}}$ for 
 $L_{\rm SN}=3\times 10^{37}{\rm erg~s^{-1}}$.
A superbubble is assumed symmetrical with respect to $x=0$, $y=0$ and
 $z=0$ planes.
Thus, 1/8 of the volume of the bubble is actually calculated.
Corresponding three symmetric boundary conditions are set for 
 $x=0$, $y=0$, and $z=0$ planes.
The other outer boundary conditions are set as free boundaries.

\section{Numerical Results}

\subsection{Model with uniform magnetic fields}

First we compare two models with $L_{\rm SN} < L_{\rm crit}$ and
 $L_{\rm SN} > L_{\rm crit}$.
Next, the effect of the distribution of magnetic field strength is studied.
In Model A, uniform magnetic fields are assumed.
Figure 1 shows the cross-cuts of the superbubble at the age of 10.4 Myr
 after the first SN exploded.
A {\em non-magnetic} superbubble consists of two parts, that is, an inner
 cavity which is occupied with a hot gas ejected by SN explosions
 and a shell which consists of ISM compressed by the effect of inner 
 high-pressured cavity.
This basic structure is common in magnetized superbubbles.
Figure 1 (a) indicates that 
 the bubble elongates to the direction of global magnetic fields
 on the mid-plane of the galactic disk ($z=0$ plane).
Hot low-density gas ejected from SNe fills a hot inner cavity
 which extends up to $x_c \la 220$ pc (near the $x-$axis) and 
 $y_c \la 150$ pc (near the $y-$axis).
Outside the hot cavity an outward-facing fast-magnetosonic shock front is
 propagating,
 which is traced as a series of sharp bends of the magnetic fields.
Since the magnetic fields bend as they reach the front after passing the
 front, this is a fast-mode shock front.
The structure perpendicular to the disk is seen in  Figs 1(b) and 1(c).
In Fig.1(b), iso-density lines (solid lines) and 
 contour lines of magnetic flux density ($B_x$) in the direction perpendicular 
 to the plot plane (e.g., $x=0$ plane) are shown. 
The outer MHD shock front is clearly traced in these figures both by 
 the distribution of the 
 magnetic flux density (dotted lines in Fig.1(b))
 and bends of the iso-density contour lines (solid lines in Figs 1(b) and (c)).
It should be noticed that the shock front is detached from the hot cavity
 and it runs  much faster than the contact surface between the ejected 
 matter and accumulated ISM.
This indicates that a {\em thick shell} is formed between the contact
 surface and the fast-mode shock front.
This is much different from the non-magnetic superbubble (Tomisaka \& Ikeuchi
 1986; MacLow et al. 1989), in which a hot cavity is surrounded by
 a cooled {\em thin} shell.
This is understood as follows:
 since the magnetic force directs perpendicular to the fields,
 compression perpendicular to the field is blocked and 
 the shell becomes much thicker than that of a non-magnetic superbubble.

In Fig. 2, physical quantities are plotted along the axes.
Temperature distribution (dotted lines) clearly indicates the boundary between 
 a hot cavity with $T\sim 2\times 10^6$K and a shell with $T \la 10^4$K.
Since the hot cavity is filled with a matter ejected by SN explosions,
 in the cavity there is no magnetic field  
 which has its origin in the  interstellar space.
It is shown that on the $x-$axis a thin shell is formed around $x\sim 250$ pc.
A magnetosonic wave front is propagating at $y_s\simeq$ 350 pc on the $y-$axis
 and at $z_s \simeq$ 600 pc on the $z-$axis, while the size of the hot
 cavity is no more than $y_c \simeq$ 150 pc and $z_c\simeq$ 200 pc.
In a region $y_c < y < y_s$ and $z_c < z < z_s$, a thick shell is expanding.
After 15 Myr, another magnetosonic wave is detached from the 
 the contact surface and propagates upwardly inside the thick shell.

In $t=22$ Myr, the hot cavity expands up to $x_c \la$ 420 pc,
 $y_c \la$ 160 pc, and $z_c \la$ 330 pc (Figs 3 and 4).
The elongated structure to the $x-$direction is characteristic of the
 magnetized superbubble.
The magnetosonic wave front seen in Fig.1(b) has propagated away
 from the frame of Fig.3(b).  
Contraction to the contact surface from outside occurs near the mid-plane
 of the disk [$v_x$ is negative in the region of
 $450{\rm pc} \alt x \alt 650{\rm pc}$ in Fig.4(a)
 and $v_y$ is negative in the region of $150{\rm pc} \alt y \alt 250{\rm pc}$ in
 Fig.4(b)].
Since the temperature of this region is lower than the average of the shell,
 this contraction seems to be driven by an effective cooling near the contact
 surface.  

Although the fast-mode magnetosonic wave has propagated away out of the
 panel of Fig.3 ($z > 1.3$ kpc),
 a hot cavity is confined below $z \la 320$ pc. 
Expansion of contact surface in each direction, $x_c$, $y_c$, 
 and $z_c$, is plotted against the age in Fig.5.
Compared with Model Q (non-magnetic case), it is shown that
 in the $x-$direction the hot cavity expands faster than Model Q.
This means that the pressure in the cavity which drives the outer shell 
 works efficiently near the $x-$direction, since there is no magnetic
 force in this direction.  
In contrast, in the $y-$direction the expansion is suppressed due
 to the magnetic tension force.
The expansion in the $y-$direction ends before $t \la 17$Myr,
 while the expansion parallel to the global magnetic fields 
 continues as long as the simulation continues for $t=22$ Myr.
Contraction in the $y-$direction seems to be driven mainly by
 the Lorentz force which works to compress the hot cavity in the 
 negative $y-$direction.
The expansion speed in $z-$direction is still positive at the age of 
 $t=22$Myr but it is heavily decelerated by the effect of magnetic tension
 force working downward.
The height of the non-magnetic superbubble in Fig.5 reaches $z_c\simeq 770$
 pc in $t=22$Myr and it indicates only a small deceleration.
While, a superbubble with the same energy release rate and age but
 formed in a magnetized ISM with $B_0=5\mu$G is almost confined to the
 galactic disk $z_c\simeq 300$ pc.
From this model, it is shown that the hot gas contained in the cavity does not
 break through the disk and is confined in the disk.
Although the magnetic force works downwardly, we found no down-flow
 in the thick shell. 
 
\subsection{The effect of the supernova rates}

In this section, we compare the models with different supernova rates.
That assumed in Model B ($L_{\rm SN}=3\times 10^{38}{\rm erg~s^{-1}}$)
 is 10 times larger than that of Model A. 
In Fig. 6 the snapshot at $t=22$Myr is plotted. 
Compared with Figs 3 and 4 ($t=22$Myr),
 it is clear that the superbubble expands much faster than Model A. 
The top of the hot cavity reaches $z_c\simeq 850$pc, while in Model A
 it is only 330 pc.
Figure 6(a) indicates that cross-cut view along the surface of $z=0$ is
 rounder than  that of Fig.3(a).
This seems to come from  a fact that the magnetic force
 becomes relatively
 unimportant compared with the ram pressure driven by the sequential 
 explosions in this case.
Comparing Figs 6(b) ($x=0$ plane) and 6(c) ($y=0$ plane), it is shown that the
 bubble is expanding preferentially in the direction parallel to the 
 magnetic fields. 
Since the bubble is confined by the magnetic fields in $y-$ and $z-$directions,
 the bubble expands along the magnetic field lines after reaching
 the halo ($400 {\rm pc}\alt z \alt 600{\rm pc}$).

Figure 6(b) clearly indicates the bubble consists of two blobs
 which are connected vertically with each other.
This is also seen in other models.
Although the bubble is almost spherical initially,
 the expansion of the top of the shell is accelerated progressively
 due to a steep vertical density gradient and the shape is deformed.

Shown in Fig.5(c),  the expansion in the $z-$direction, ${z_c}$,
 is decelerated due to the magnetic force but $dz_c/dt$ is still positive.
Although the expansion speed in the $y-$direction becomes negative
 (contracting) in a later phase of Model A, this model shows a positive
 velocity (expanding) throughout the evolution.
This is also explained by the fact that the ram pressure force is relatively
 important against the magnetic tension which works to decelerate
 the expansion in the $y-$direction.
In the $x-$direction, expansion continues without any deceleration.

Summarizing the results for Models A and B,
 the expansion in the $z-$direction is more or less suppressed
 by the uniformly distributed magnetic fields running in the $x-$direction.
The hot gas seems to be confined by a uniform magnetic field
 with the strength of $B_0=5\mu$G to the extent of
 $|z| \la 400$pc (Model A) and $|z| \la 1$kpc (Model B),
 when the scale-height of the distribution of magnetic field strength
 is much larger than that of the density as Models A and B.
These models correspond to the infinite $H_B$.
In the following section, we focus on the case with a finite $H_B$.

\subsection{The effect of distribution of the magnetic fields}

In this section,
 we focus on the effect of {\em distribution} of the magnetic fields.
In Model C, we assume that the strength of magnetic fields
 decreases from disk to halo.
The ratio of the thermal energy density to the magnetic one is assumed 
 constant.
That is, since the gas temperature is constant, 
 the magnetic field strength changes according to the density as
\begin{equation}
  B_x(z)=B_0 \left(\frac{\rho(z)}{\rho_0}\right)^{1/2},
\end{equation}
where $B_0$ and $\rho_0$ represent, respectively, the magnetic field
 strength and the gas density on the mid-plane of the galactic disk ($z=0$).
This gives a scale-height measured by the magnetic energy
\begin{equation}
  H_B=\frac{1}{B_0^2}\int_0^\infty B_x^2(z)dz,
\end{equation}
equal to the density-scale height $H$.
The density distribution is again assumed as a model
 proposed by Dickey and Lockman (1990).
Then, since the gas is supported by the magnetic pressure in contrast to Models
 A and B, the corresponding gravity becomes weak.              
Model C assumes the same mechanical luminosity $L_{\rm SN}$ as Model A.

The snapshots at the ages of $t=10$Myr (panels a-c),
 $t=22.8$Myr (panels d-f),
 and  $t=30$Myr (panels g-i) are plotted in Fig. 7.
As shown in Fig.7(b), the hot cavity consists of two blobs similar
 to Figs 3(b) and 6.
This panel captures a structure just after a top of the spherical
 shell is deformed and hot gas outflowing through the hole is re-expanding
 into the halo region.
The reason why the hole is made seems to an instability which destructs
 the shell.

The pressure in a hot cavity decreases with time as
 from $p\sim 6000\,{\rm K~cm^{-3}}$ at the age of $t=10$Myr
 to $p\sim 700\,{\rm K~cm^{-3}}$ at $t=30$Myr.
The volume of the hot cavity increases more than 20 times in this time span.
If the bubble were to expand exactly in an adiabatic fashion,
 the pressure would  decreases much, $\sim 1/150$.
However, it is certain that a rapid expansion into the low-density
 halo reduces the pressure rapidly after $t=10$ Myr. 

In this model, as increasing the height, the effect of magnetic fields
 decreases.
Thus, after the bubble escapes from the disk,
 it becomes hard to confine the hot gas with the magnetic fields. 
Comparing with Fig.3, even with the same mechanical luminosity as
 $3\times 10^{37}{\rm erg~s^{-1}}$, it is evident that the superbubble
 is expanding faster than that of Model A.

Comparing Fig.7(d) with Fig.3(a), 
 as for the mid-plane ($z=0$) the area occupied with a hot gas  
 is smaller than that of Model A.
Contraction driven by a magnetic tension occurs also in this model.
In Fig.5(b), the time evolution of $y_c$ is illustrated.
This shows that the maximum size of the superbubble on the mid-plane
 perpendicular to the magnetic field, $y_c$, {\em decreases} after 
 $t\simeq 10$Myr (Figs 7(a)-(c)).
This occurs after the pressure of the cavity decreases due to a 
 rapid expansion of the bubble into the halo. 
The shell moves reversely in the $y-$direction by the effect of the magnetic
 tension.
In the course of the contraction, a part of the shell in $y > 0$
 and that in $y < 0$ merge together.
At that time, the size in the direction parallel to the magnetic field,
 $x_c$, decreases abruptly ($t\simeq 22$Myr).

Figures 7(b), 7(e), and 7(h) indicate
 that the hot cavity is surrounded by a thick shell
 viewing from the direction of the magnetic field.
At the age of $t=30$ Myr, 
 in a region near the $z-$axis ($z\simeq 200-300$ pc),
 the hot matter ejected with supernova explosions flows through a 
 narrow channel like a throat.
A long sheet is extending from the throat, which was originally 
 a shell on the top of the bubble.
Figures 7(c), 7(f), and 7(i) shows that the shell 
 propagating along the magnetic fields  is also thick.
The interface between the shell and the hot interior is clearly seen
 as a sharp density discontinuity.

It is concluded that in a galactic disk
 whose magnetic scale-height, $H_B$,
 is as small as that of the density, $H$, 
 a superbubble evolves like a bubble in a water.
That is, a low density bubble rises by the effect of buoyancy.

\subsection{Low luminosity bubble}

In the preceding subsection, it is shown that 
 an outflow of hot gas is realized 
 in a galactic disk with a small magnetic scale-height.
Then, does a less luminous bubble make the hot gas flow into the halo
 even from the same initial conditions as Model C?
In Model E, a poor OB association is assumed with a 
  mechanical luminosity of $L_{\rm SN}=10^{37}{\rm erg\,s^{-1}}$.
The expansion law for this model is also plotted in Fig.5.
Owing to a low luminosity, the expansion is slower than Model C.
 
Figure 8 shows the snapshot at the age of $t=40$Myr.
As shown in Fig.5, the snapshot corresponds to the structure
 after the contraction occurs due to the magnetic tension.
Comparing with Model C, it is evident that the volume occupied with
 a hot cavity is small.
A fraction of the hot cavity to the entire volume inside the magnetosonic
 shock wave front is also much smaller than that of Model C.
From this figure, since the volume of hot gas is small and
 the height of the top $z_c$ is no more than $\sim 500$ pc,
 it is concluded that less luminous bubbles do not play  important roles
 as an origin of the interstellar hot gas of $T \la 10^6$K.

\section{Discussion}
\subsection{Blow-out or confinement?}

As shown in the preceding section,
 although in Model B the mechanical luminosity is much larger than the critical
 luminosity of equation(\ref{eqn:Lcr0}), the vertical expansion is much 
 decelerated by the effect of magnetic fields.
This shows that $L_{\rm SN} < L_{\rm crit}$ may be only a sufficient
 condition for confinement of the superbubble.

The expansion of a spherical shell driven by a pressure in the hot cavity $p$
is formulated as follows: 
\begin{equation}
\frac{dMv}{dt}=4\pi\, R^2 (p-p_{\rm out}),
\end{equation}
\begin{equation}
\frac{dM}{dt}=4\pi\, R^2\, v\, \rho_{\rm out}
\end{equation}
\begin{equation}
\frac{dR}{dt}=v,
\end{equation}
\begin{equation}
\frac{dE}{dt}=L_{\rm SN}-4\pi R^2\, p\, v,
\end{equation}
where, $M$, $R$, $v$, $p_{\rm out}$, $\rho_{\rm out}$ and 
 $L_{\rm SN}$ represent, respectively, 
 the mass of the shell, the radius and velocity of the shell, 
 the interstellar pressure and its density and the energy release rate from
 an OB association.
These equations are, respectively, the equation of motion,
 the mass conservation, the relation of a size $R$ to a velocity $v$, and
 the first law of thermal physics. 
If we assume $p=(2/3)\times E /(4\pi R^3/3)$, these four equations can be solved 
numerically.
Figure 9 shows a resultant expansion law of a spherical superbubble with 
 $L_{\rm SN}=3\times 10^{37}{\rm erg~s^{-1}}$ in a {\em uniform 
 interstellar medium} of $n_0=0.3{\rm cm^{-3}}$.
Each curve corresponds to different external pressures as
 $p_{\rm out}=1.7\times 10^{-12}{\rm erg~cm^{-3}}$ (solid line),
 $p_{\rm out}=1\times 10^{-12}{\rm erg~cm^{-3}}$ (dotted line),
 and $p_{\rm out}=0$ (dashed line).
Weaver et al.'s (1977) solution, equation (1),
 agrees with the curve of $p_{\rm out}=0$.
This figure shows that  the interstellar pressure plays an important role
 especially in the late phase of the evolution.
This is understood as follows: 
 in the late phase of the superbubble 
 the difference between the internal pressure
 and the outer one is small and this small difference drives the 
 shell further. 
Therefore, the critical luminosity 
 would be  underestimated 
 if we use a solution without taking  the outer pressure into account.
The shell of a superbubble 
 continues to expand as long as the energy ejection
 continues, while a supernova remnant stops its expansion after
 $p_{\rm out}=p$.
Thus, exactly speaking, the superbubble is never confined as long as
 the OB association is alive. 
However, a slow expansion driven by a small pressure difference as
 $p-p_{\rm out}$ is considered as a signature of the confinement.

In this figure, we also plot an equivalent radius, which is defined
 using the volume occupied with a hot matter ($V_{\rm hot}$)
 as 
\begin{equation}
R_{\rm equiv}\equiv \left(\frac{3V_{\rm hot}}{4\pi}\right)^{1/3}, 
\label{Requiv}
\end{equation}
 for Models A and C.
The equivalent radius for Model A in which the bubble is almost confined 
 in the galactic gaseous disk shows a similar expansion law to that obtained
 by a thin-shell model (a solid curve). 
In contrast, that of Model C indicates a completely different expansion law
 such that after $t \agt 10$ Myr the equivalent radius increases rapidly
 and in $t\simeq 35$Myr $R_{\rm equiv}$ surpasses the model with 
 $p_{\rm out}=0$.
These differences seem to come from the distribution of magnetic field 
 strength.
In Model A the total pressure (thermal plus magnetic one) is almost constant
 as the bubble expands,
 because the magnetic pressure is dominant over the thermal one and
 magnetic fields are uniform.
While, in Model C the total pressure drops according to the density  
 distribution $\rho(z)$.
This figure indicates that when the equivalent radius is well fitted
 by this thin shell model the bubble is nearly confined in the disk
 even if the gas disk has a finite scale-height.
In contrast, if hot gas is ejected from the galactic disk,
 the equivalent radius shows a more rapid expansion than
 that derived by this thin-shell model.
Values of equivalent radii at the epochs when numerical runs end are
 shown in table 1. 

\subsection{Observability}

A shear motion in the galactic rotation and rotation itself
 may play a role in the evolution of a superbubble
 (Tenorio-Tagle \& Palou\v{s} 1987; Palou\v{s} et al. 1990;  Silich 1993).
Galactic shear seems to deform the shape and the Coriolis force
 makes the shell rotate.
The characteristic time-scales of the rotation, $\tau_R$, and the shear,
 $\tau_S$ are estimated respectively
 as 
\begin{equation}
\tau_{\rm R} \sim 1/\Omega_0 \sim 40{\rm Myr} 
 (\Omega_0/26{\rm km~s^{-1}~kpc^{-1}})^{-1}
\end{equation}
and
\begin{equation}
 \tau_{\rm S}\sim (l d\Omega/dR)^{-1} \sim 320 {\rm Myr}
 (l/1{\rm kpc})^{-1}(\Omega_0/ 26{\rm km~s^{-1}~kpc^{-1}})^{-1}
 (R_0/8.5{\rm kpc}),
\end{equation} 
 where $\Omega_0$, $R_0$, and $l$ are the angular speed of galactic
 rotation, distance from the galactic center, and a typical size of
 a superbubble, respectively. 
Since active  SN-explosion phase continues for $\sim 50$ Myr for
 an OB association (McCray \& Kafatos 1987), 
 in the late phase of $\tau_{\rm R}\la t \la 50$Myr, the effect of the Coriolis
 force seems to appear as a deformation force of the shell.  
The $\alpha\omega$-dynamo mechanism driven by superbubble was 
 studied recently by Ferri\`{e}re (1992).

The  galactic rotation has a little effect on the evolution of a superbubble.
Thus, if shells or holes observed in external galaxies are elongated after  
 their inclinations are corrected, their direction seems to indicate that of
 the magnetic fields. 
There have been listed 141 HI holes in M31 by
 Brinks \& Bajaja (1986).
These holes are observed, more or less, as elliptical.
Particularly, the holes found near the major axis of M31 are
 important to determine the physical shape of the holes.
Many of these HI holes  have their major axes 
 perpendicular to the galaxy's major axis.
Since this is not explained by projection due to the inclination of M31 
 ($i=77\deg$),
 these seem to have physically a shape elongated along the azimuthal 
 direction of the galaxy.
This is not inconsistent with  observations indicating that 
 a global pattern of the magnetic field is ring-like in M31, which is measured
 by  radio  linear polarization observations 
(for a review, see Sofue, Fujimoto \& Wielebinski 1986), 
 in other words, magnetic field lines run in the azimuth direction
 in M31.

\subsection{Porosity}

If  hot gas contained in superbubbles occupies a large volume of the 
 galactic disk, a picture of the interstellar medium should be changed
 (McKee \& Ostriker 1977).
The fraction of areas covered by superbubbles younger than $\tau_{\rm active}$ 
 is estimated with a quantity called as two-dimensional porosity which
 is defined as 
\begin{equation}
 Q(t < \tau_{\rm active}) \equiv r_{\rm OB}\int_0^{\tau_{\rm active}} S(t)dt,
\end{equation}
where $r_{\rm OB}$ is the formation rate of OB associations per unit area
 and $S(t)$ represents the area which covered by a hot cavity on the  
 mid-plane of the disk $z=0$.
This is identical with a two-dimensional porosity parameter calculated
 by  Heiles (1990).
He estimated galaxy-wide average of two-dimensional porosity 
 $Q_{\rm 2D}\simeq 0.30$. 
$\tau_{\rm active}$ should be chosen equal to
 the oldest age of a superbubble which contains a hot gas inside.
If we assume $\tau_{\rm active}=20$ Myr and integrate $S(t)$ for Models
 A and C ($S\equiv \pi x_c y_c$),  these two models give respectively
  $2.08\times 10^6\, {\rm pc^2~Myr}$ and $1.57\times 10^6\,{\rm pc^2~Myr}$.
We adopt the estimation of $r_{\rm OB}$ from a galactic type II supernova rate
 of
 $r_{\rm II}\sim 0.01{\rm yr^{-1}}$, that is, we assume that all type
 II SNe occur in OB associations, number of type II SNe
 in an OB association is constant irrespective of richness of association
 as $N_{\rm SN}\sim 100$ and  OB associations  are uniformly distributed 
 in the galactic disk with radius $R_{\rm gal}\simeq 10$ kpc. 
This gives an estimation of OB association formation rate as 
\begin{equation}
 r_{\rm OB}  =  \frac{r_{\rm II}}{N_{\rm SN}\,\pi\, R_{\rm gal}^2},
\end{equation}
\begin{equation}
  r_{\rm OB}     \simeq  3.2 \times 10^{-7}{\rm pc^{-2}\,Myr^{-1}}
              \left( \frac{r_{\rm II}}{0.01{\rm yr^{-1}}}\right)
              \left( \frac{N_{\rm SN}}{100} \right)^{-1}
              \left( \frac{R_{\rm gal}}{10 {\rm kpc}}\right)^{-2}. 
\end{equation}
This indicates the two-dimensional porosity to be equal to
 $Q(t < 20 {\rm Myr})\simeq 0.5-0.6$.
Thus, rather large fraction of the galactic disk, $1-\exp{(-Q)}\sim 40\%-45\%$,
 is covered by young ($t < 20$ Myr) superbubbles.
 
\section*{Acknowledgments}

The author would like to thank Dr. F. Nakamura for his 
 critical reading of an earlier version of the manuscript.
This work was supported in part by Grants-in-Aid for Scientific Research from 
 the Ministry of Education, Science, Sports and Culture(07640351, 07304025).
Numerical simulations were performed by Fujitsu VPP300/16R supercomputer 
 at the Astronomical Data Analysis Center of the
 National Astronomical Observatory, Japan.

\label{lastpage}

\clearpage
\end{document}


\clearpage

\setcounter{figure}{0}
\begin{figure*}
\plotthree{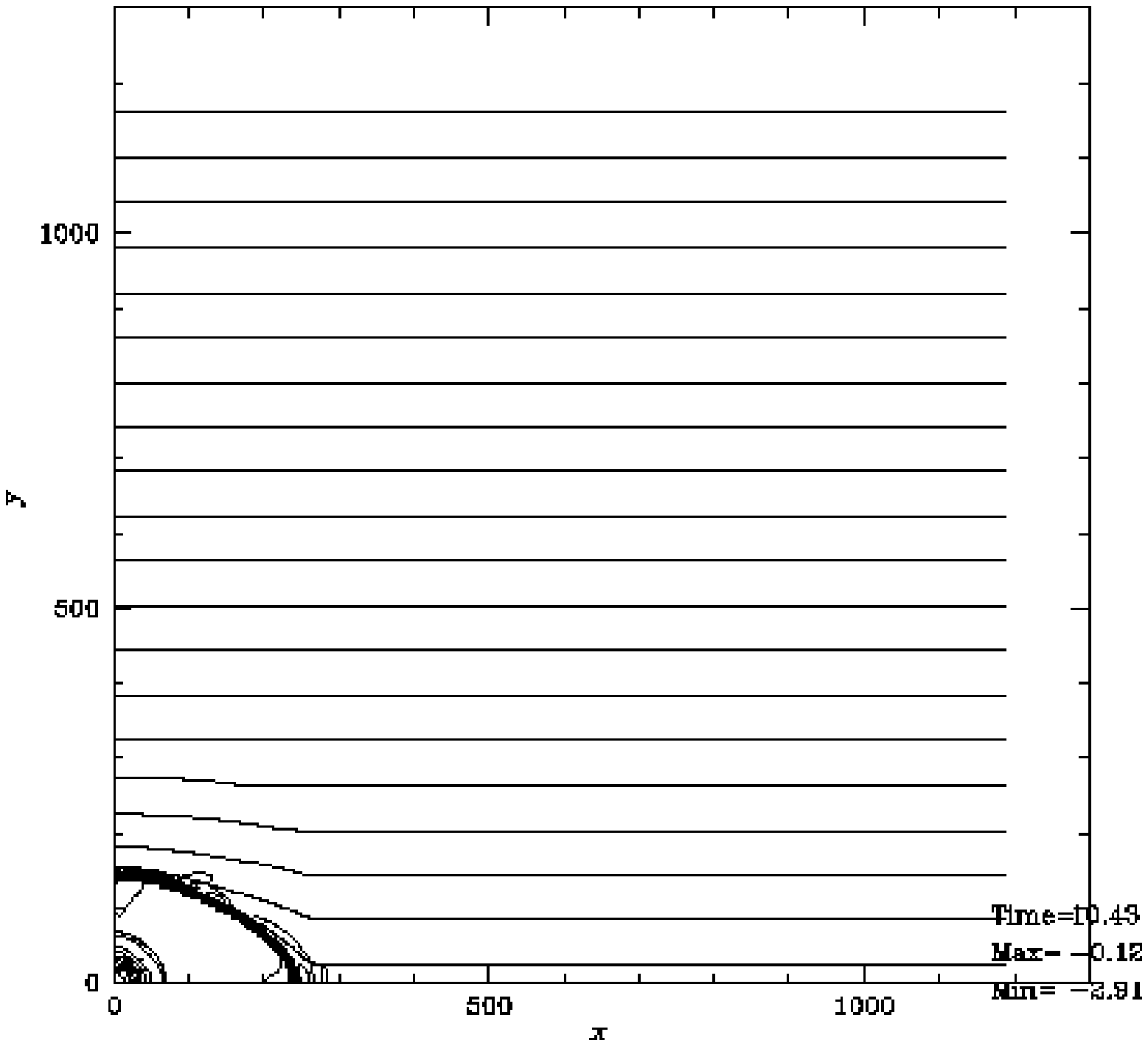}{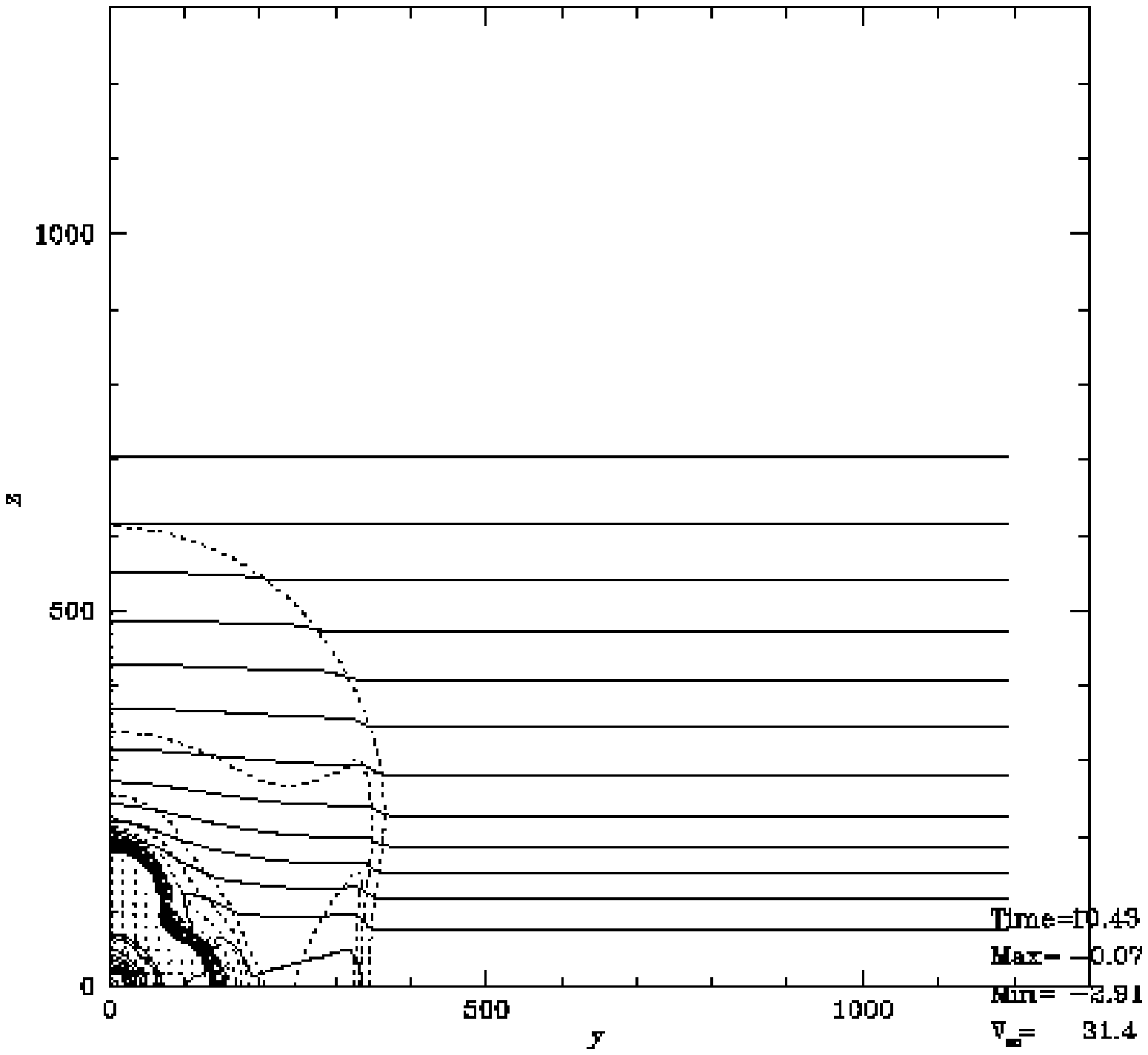}{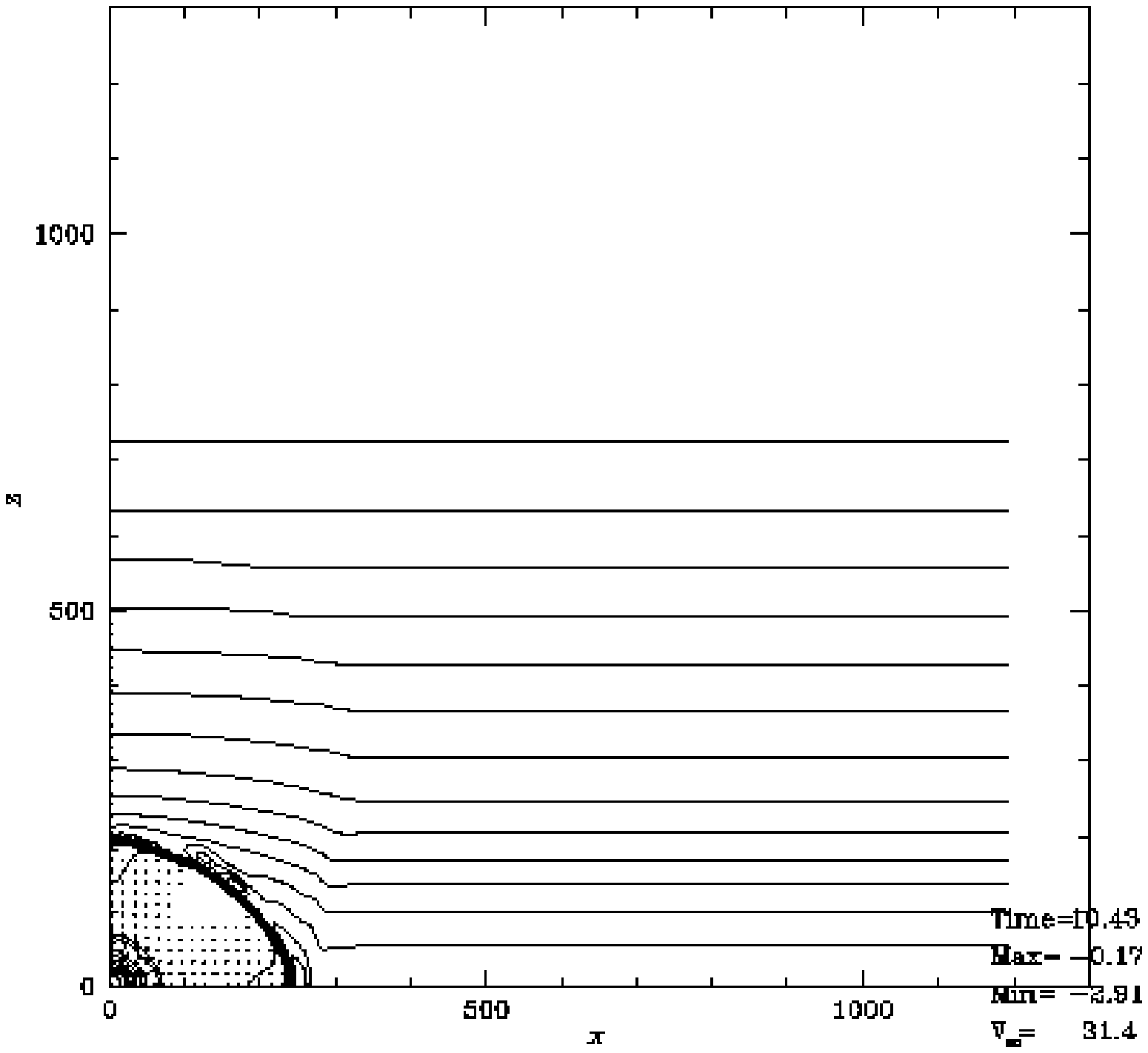}
\caption{\label{fig:1}(a): A cross-cut along the mid-plane of the disk, $z=0$,
 at $t=10.4$ Myr.
Density contour lines and magnetic field lines (running horizontally)
 are plotted.
The numbers near the bottom-right corner represent, respectively,
 the time of the snapshot, the maximum and  minimum levels of the contours,
 e.g., 21 contour lines are plotted for the values from $10^{-3.10}\rm cm^{-3}$
 to  $10^{-0.38}\rm cm^{-3}$.
 (b): The same as (a) but for the $x=0$ plane.
Density contour lines and velocity fields are plotted.
The magnetic field is running perpendicular to this plane and
 its distribution is represented by dotted contour lines. 
The maximum speed plotted in the panel 
 is also shown near the bottom-right corner.
(c): The same as (b) but for the $y=0$ plane.}
\end{figure*}

\setcounter{figure}{1}
\begin{figure*}
\plotthree{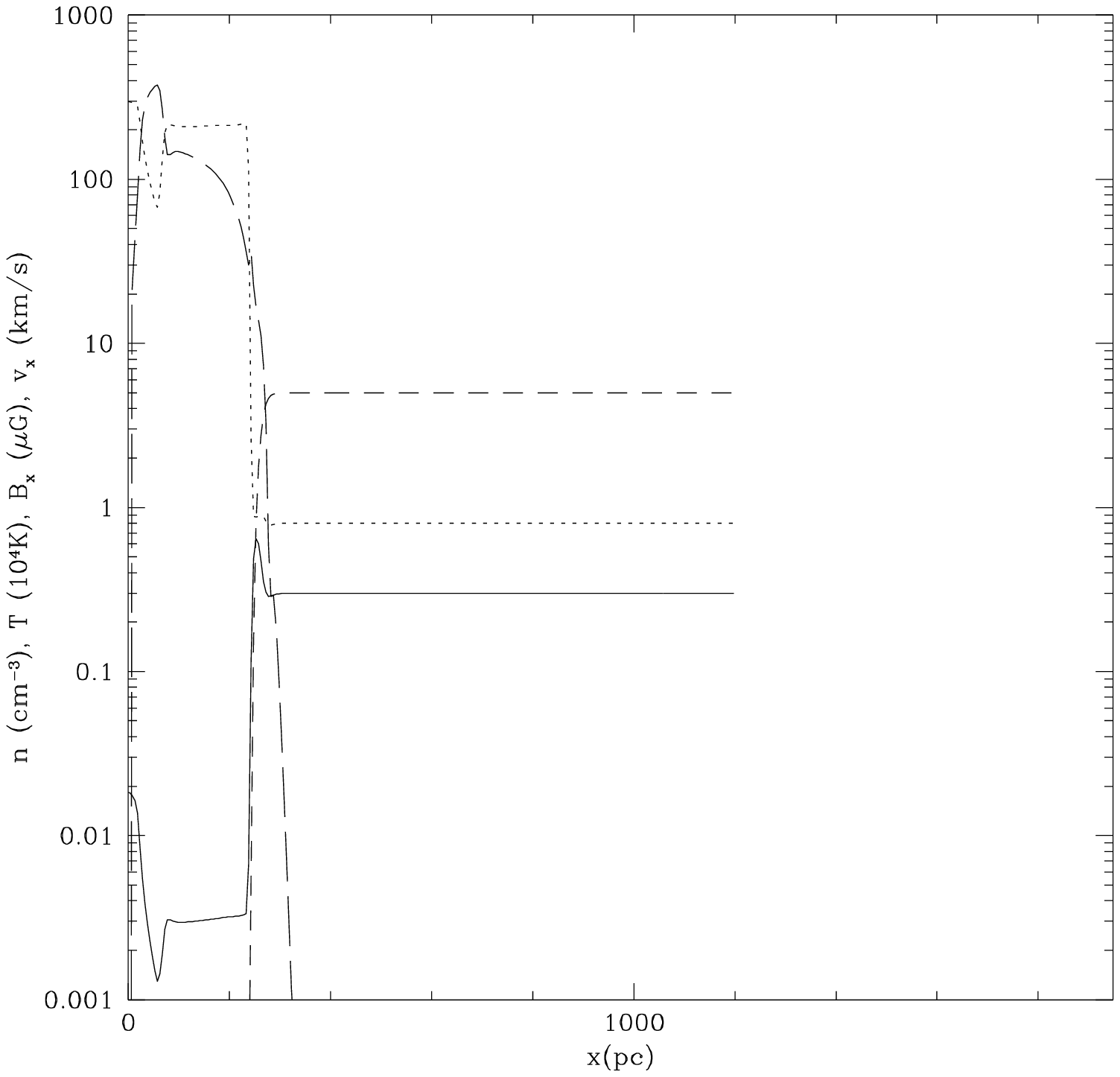}{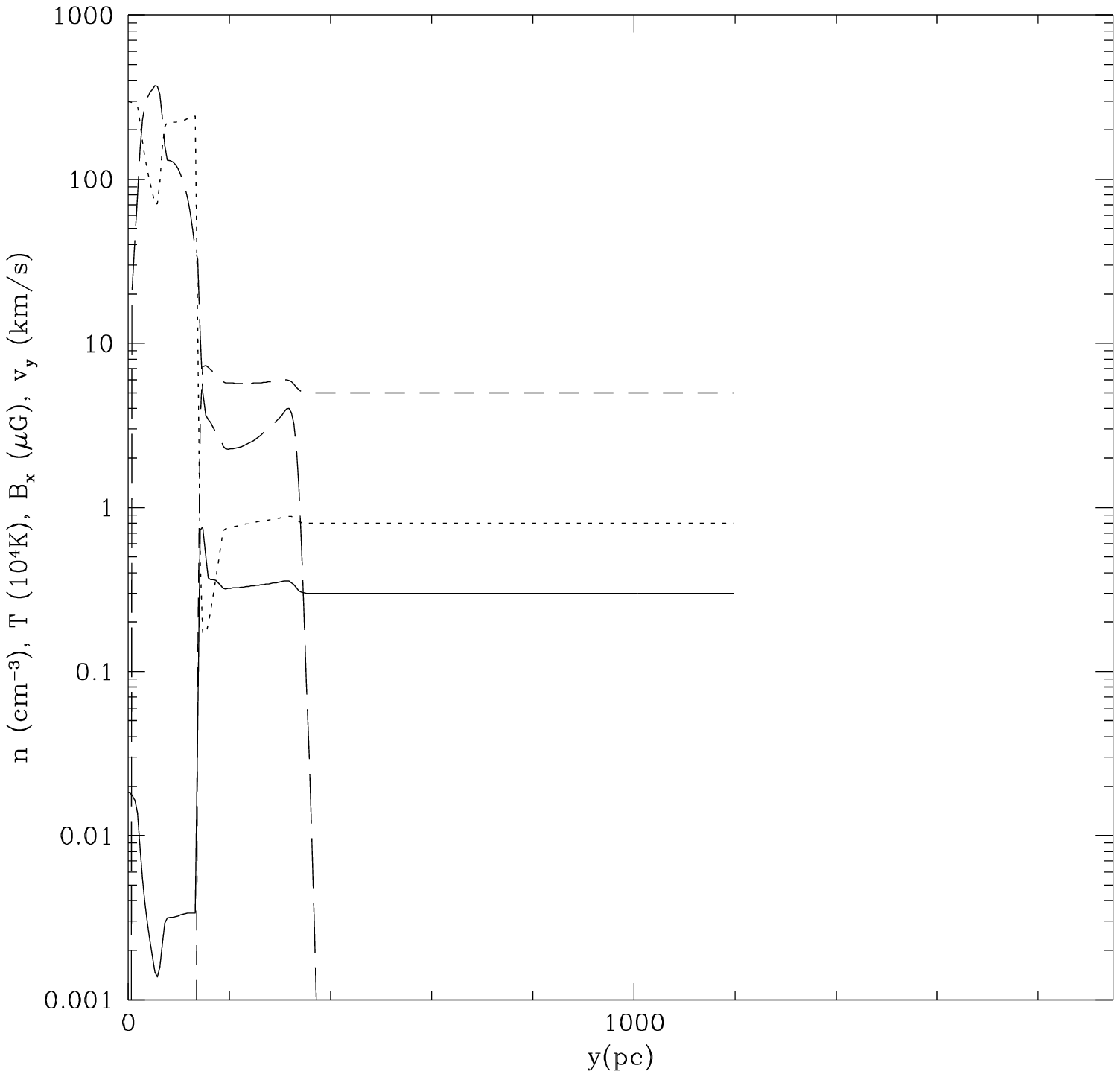}{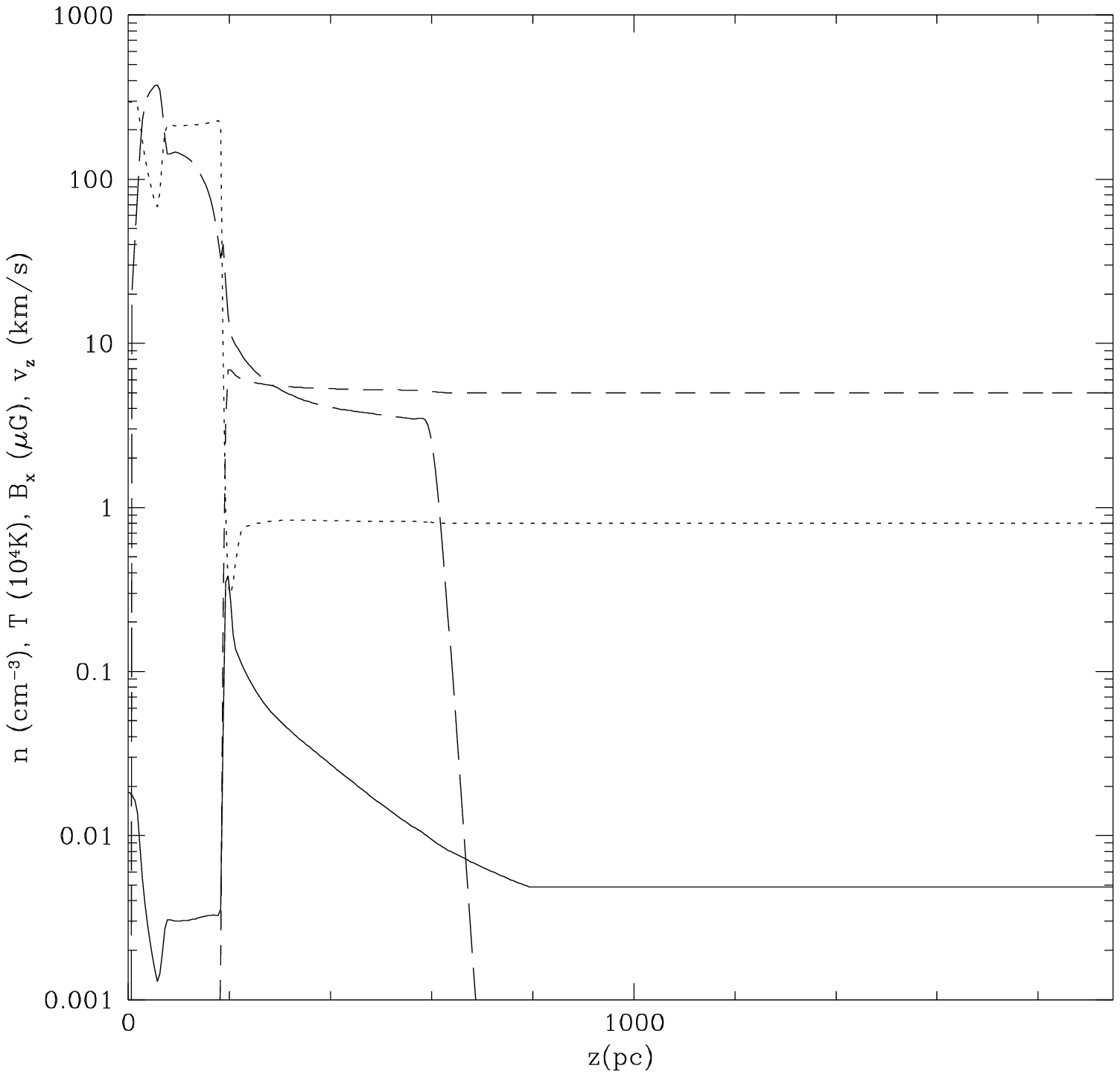}
\caption{\label{fig:2}Physical quantities are plotted along the $x-$ 
(a), $y-$ (b), and $z-$axis (c).
The density $n {\rm (cm^{-3})}$ (solid lines),
 the temperature $T {\rm (10^4 K)}$ (dotted lines),
 the magnetic flux density $B_x {\rm (\mu G)}$ (dashed lines), and
 the velocities along the respective axes $|v| {\rm (km~s^{-1}})$ (long-dashed
 lines) are plotted.
This is a snapshot at the age of $t=10.4$ Myr.}
\end{figure*}

\setcounter{figure}{2}
\begin{figure*}
\plotthree{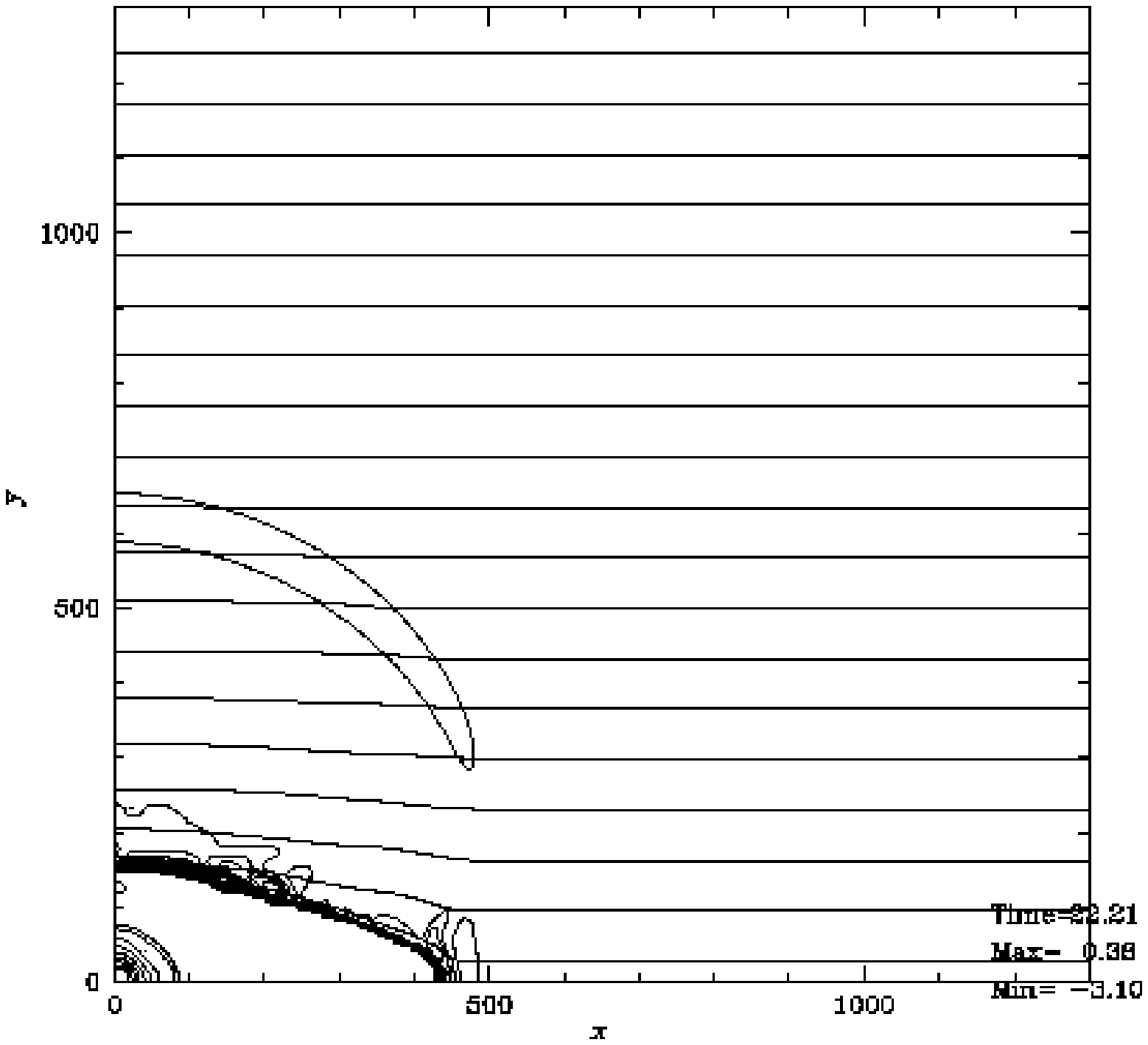}{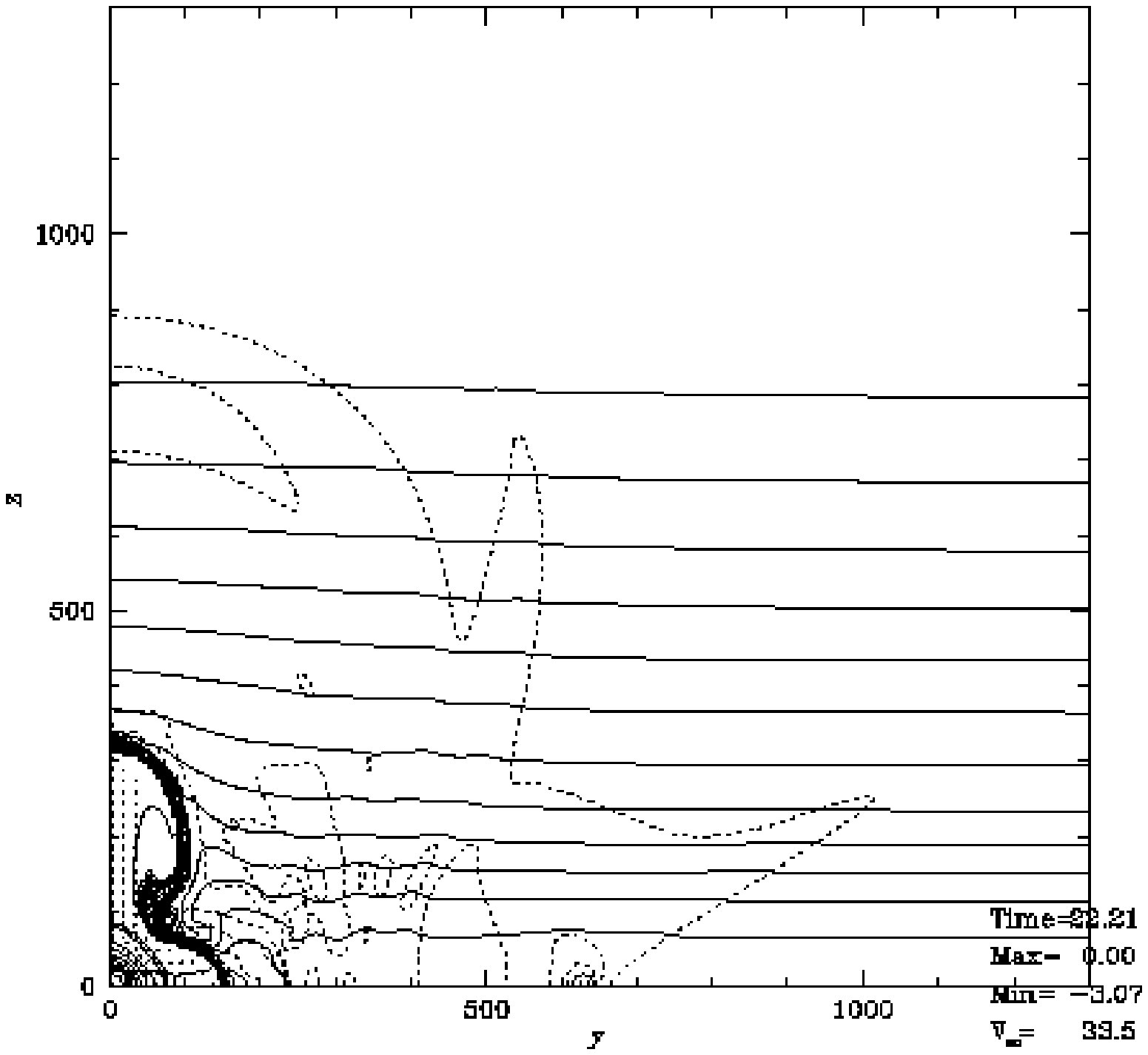}{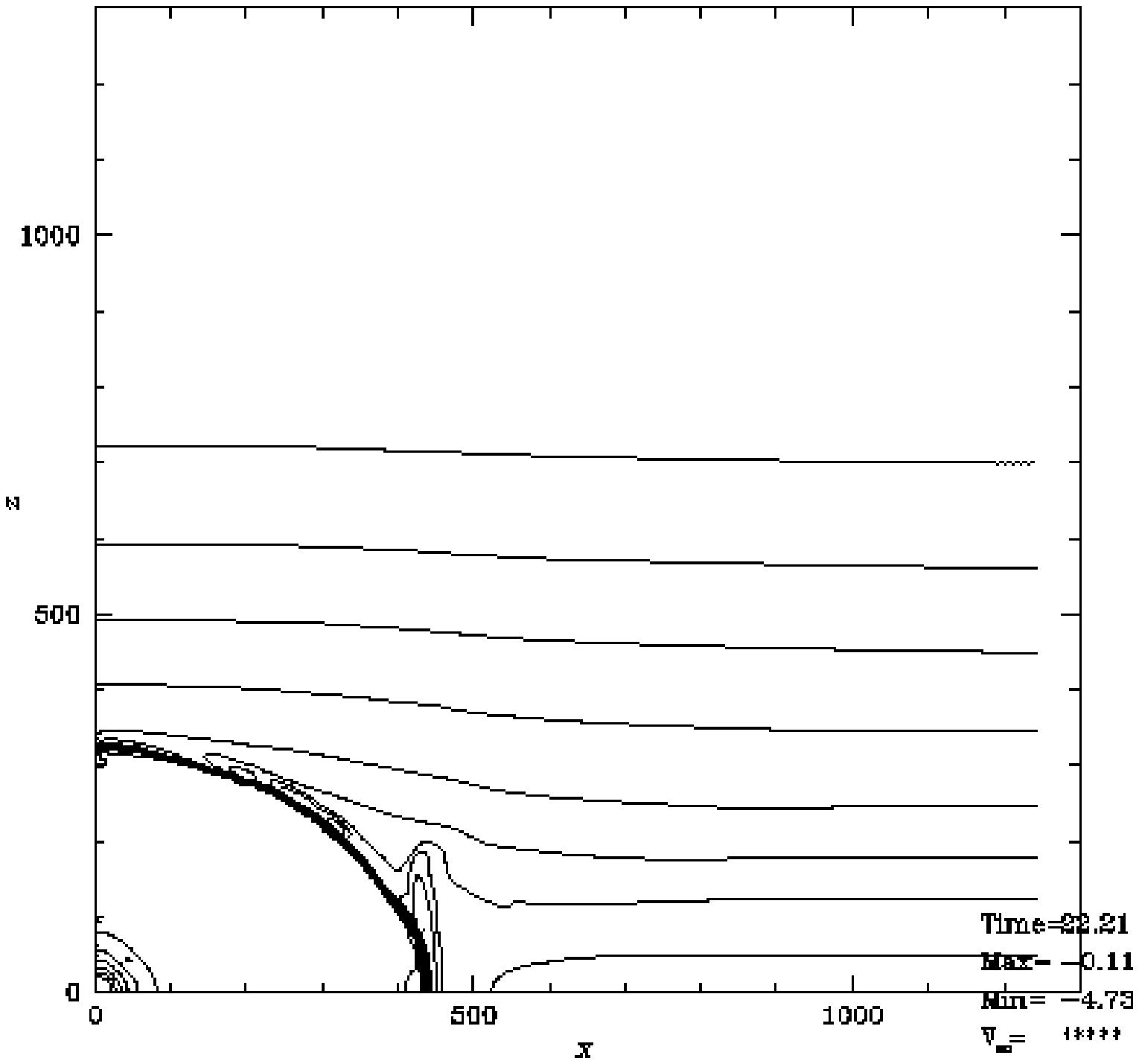}
\caption{\label{fig:3}The same as Fig.1, but for the age of $t=22$ Myr.}
\end{figure*}

\setcounter{figure}{3}
\begin{figure*}
\plotthree{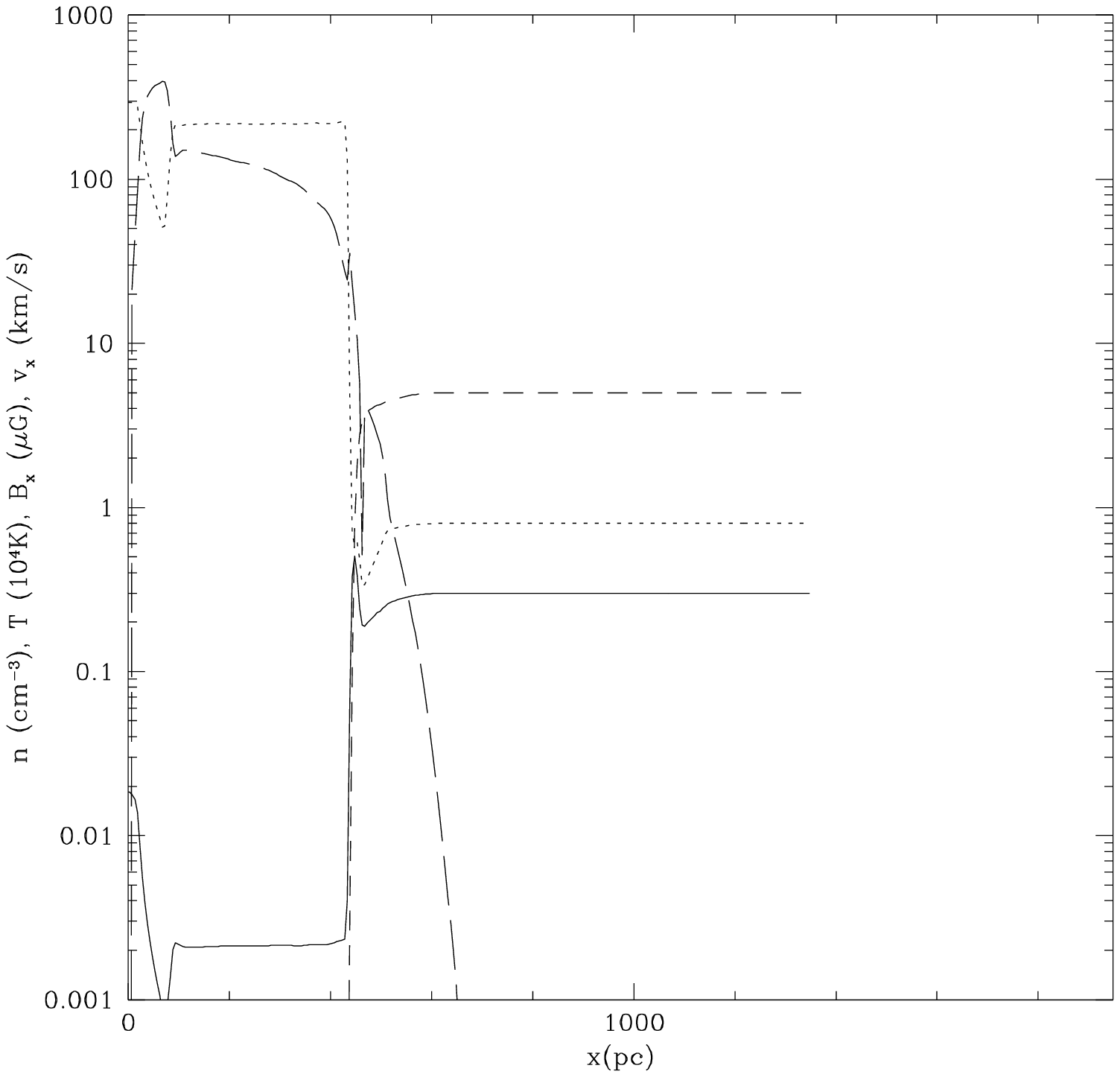}{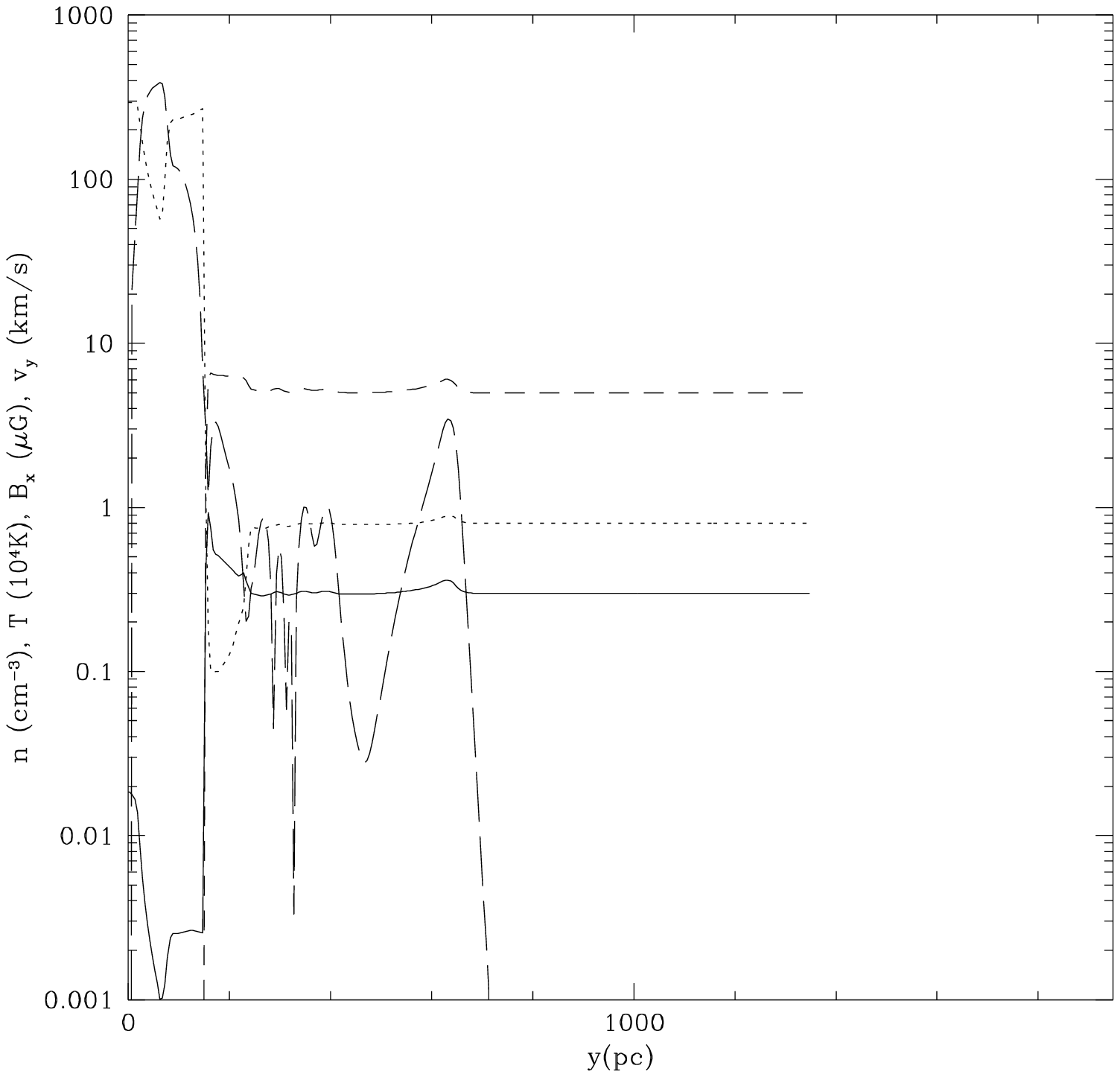}{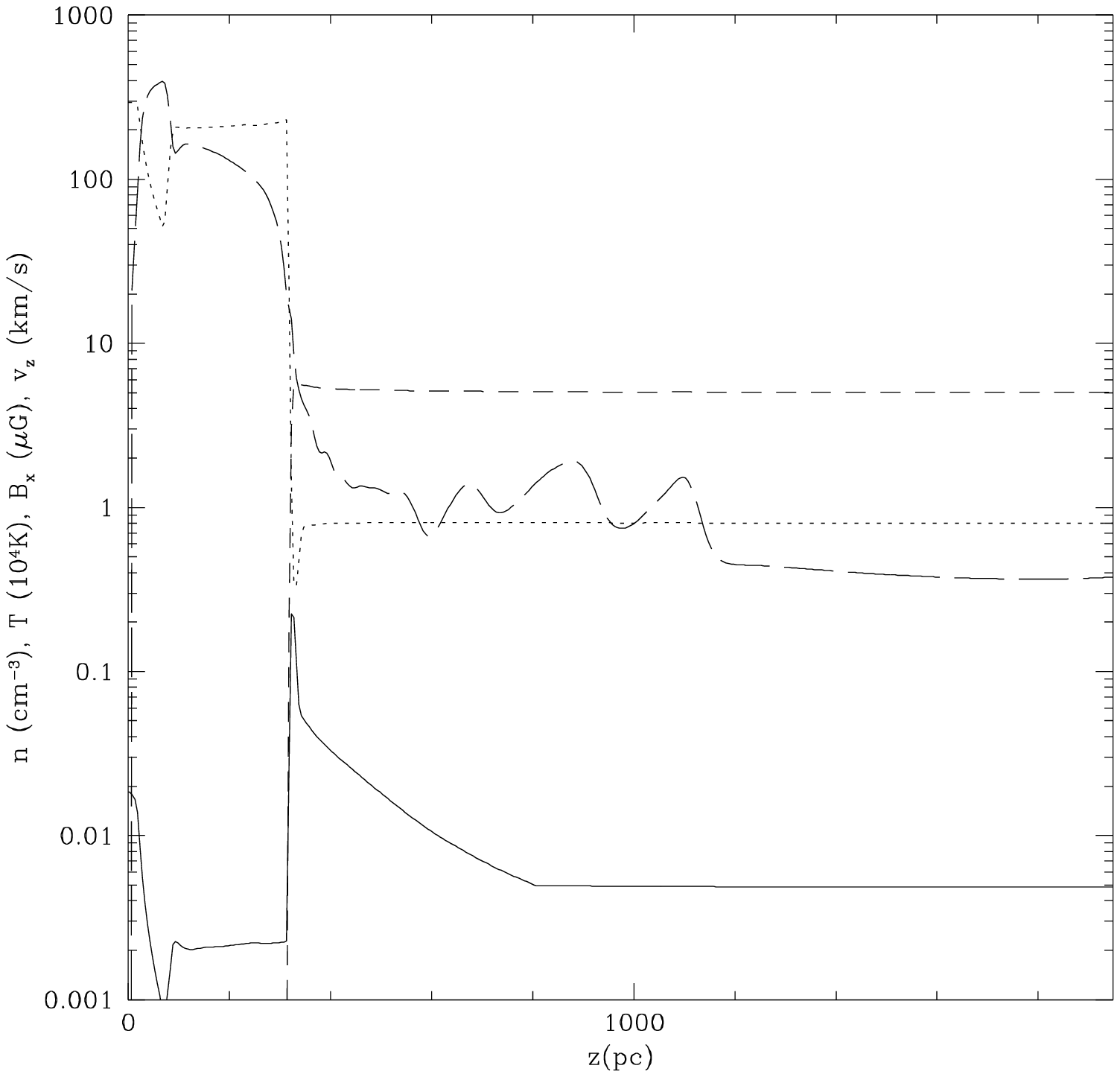}
\caption{\label{fig:4}The same as Fig.2, but for the age of $t=22$ Myr.}
\end{figure*}

\clearpage
\setcounter{figure}{4}
\begin{figure*}
\plotthree{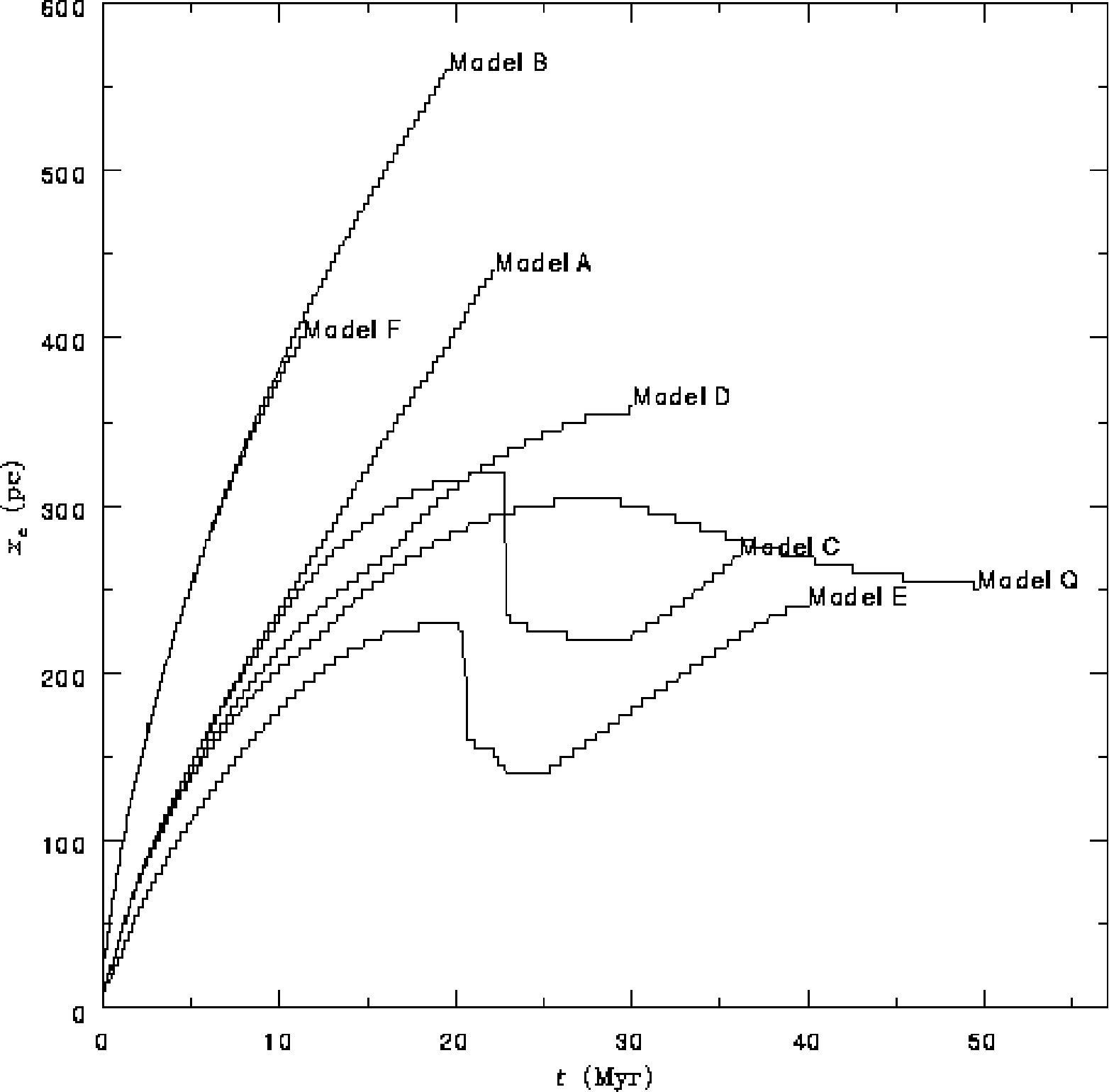}{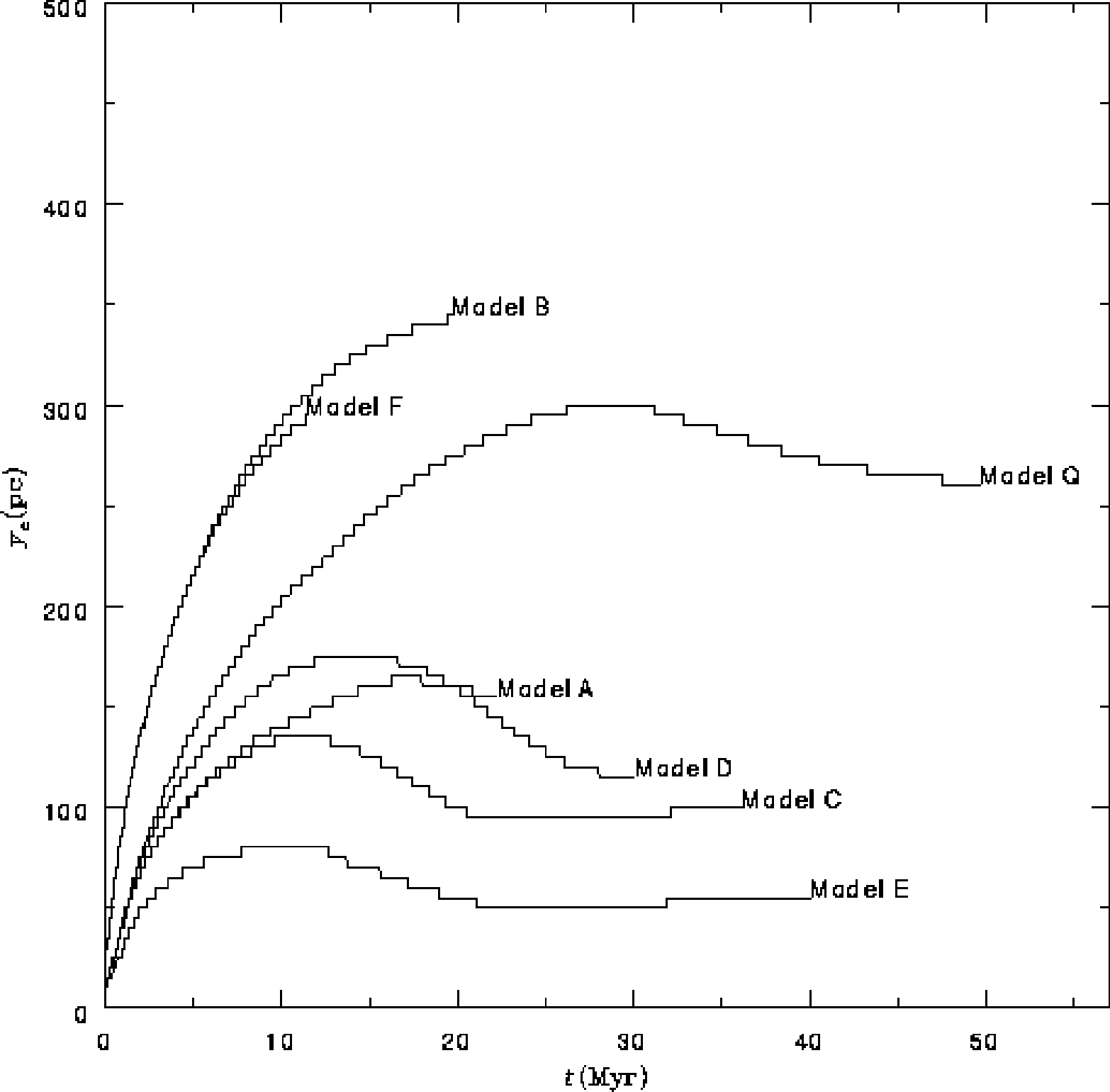}{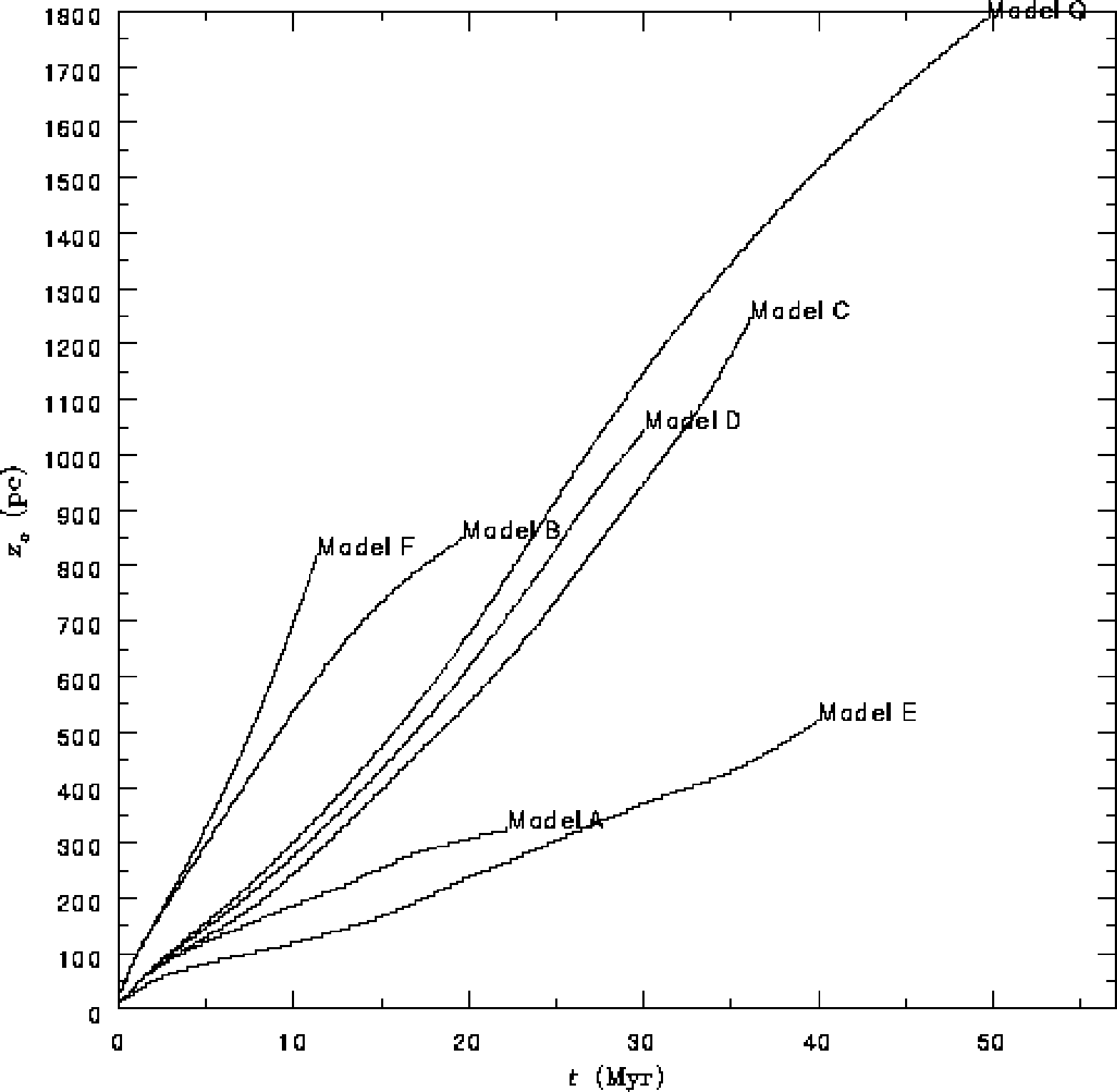}
\caption{\label{fig:5}Expansion law of a hot cavity formed in a superbubble.
The size of the hot cavity is measured with ($x_c,y_c,z_c$).
Models A and B correspond to the cases that a superbubble is formed in 
 ISM with a uniform magnetic field distribution.
In Models C, D, E, and F, the magnetic field strength was assumed 
 to decrease upwardly,
 while the Alfven speed $B/(4\pi \rho)^{1/2}$ is constant.
Model Q, non-magnetic superbubble, is also plotted.}
\end{figure*}

\setcounter{figure}{5}
\begin{figure*}
\plotthree{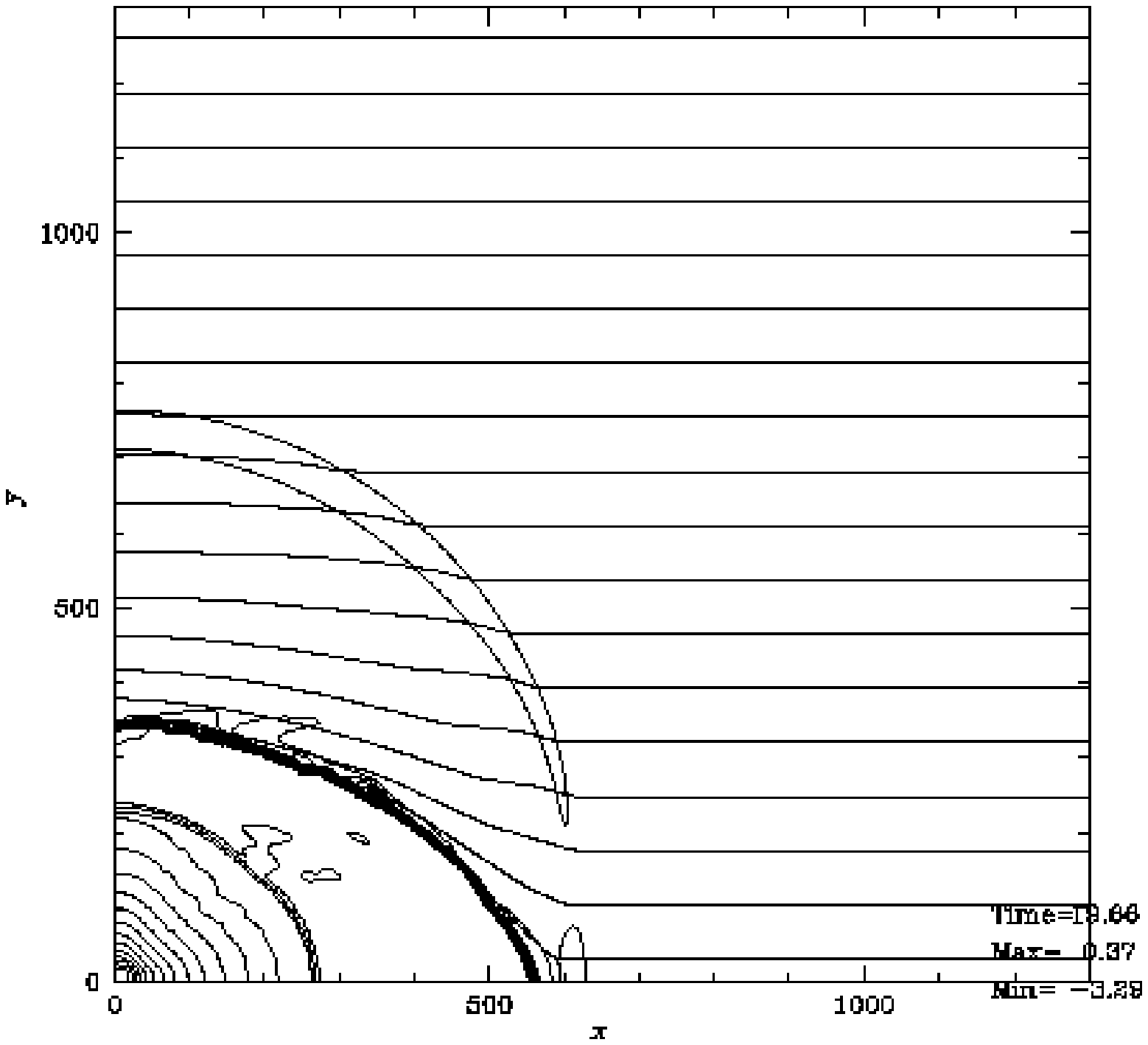}{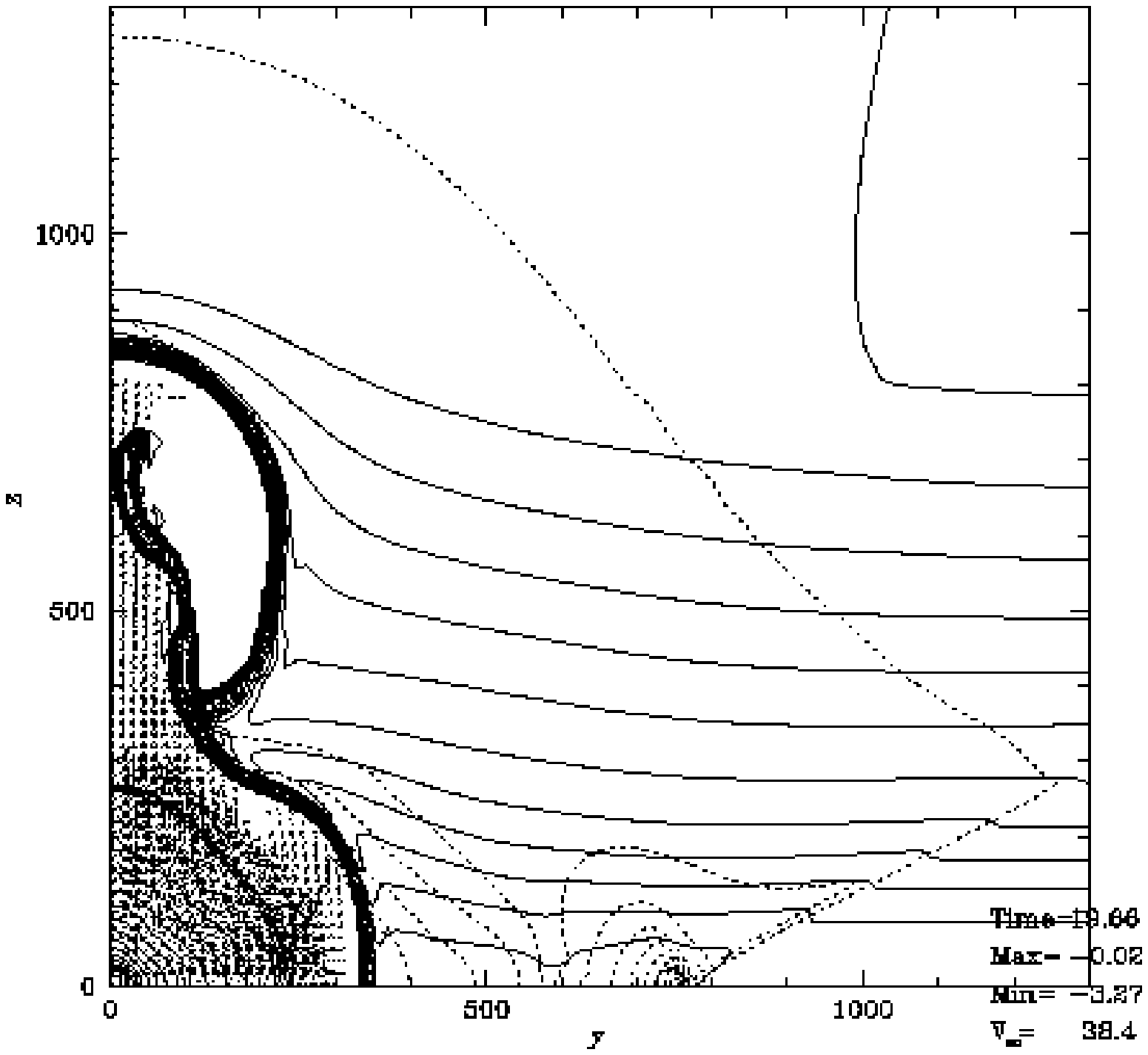}{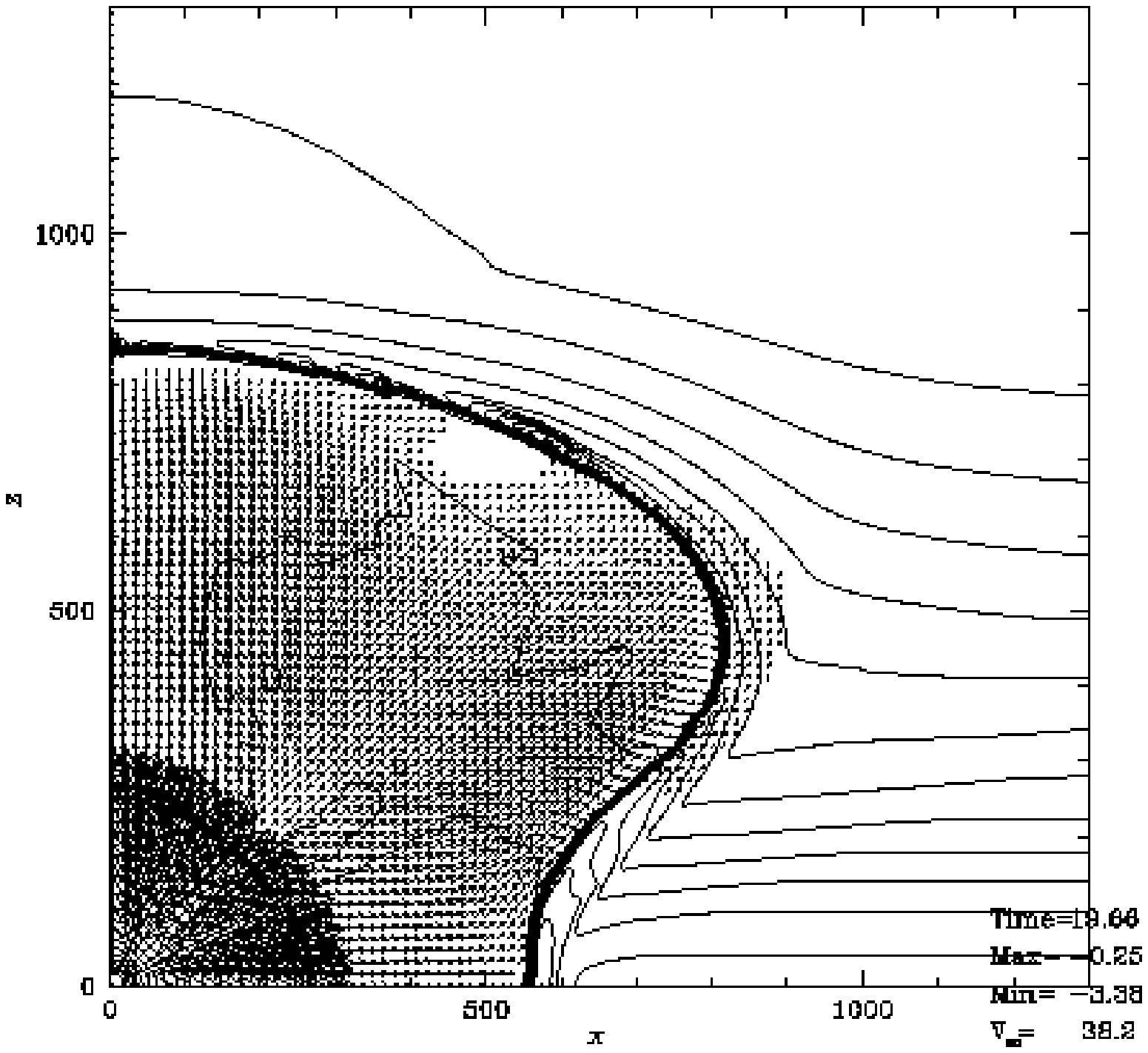}
\caption{A superbubble driven by an active OB association 
 with a luminosity 10 times larger than the
 previous model. 
This is a snapshot at the age of $t=20$ Myr.}
\end{figure*}

\clearpage
\setcounter{figure}{6}
\begin{figure*}
\plotthree{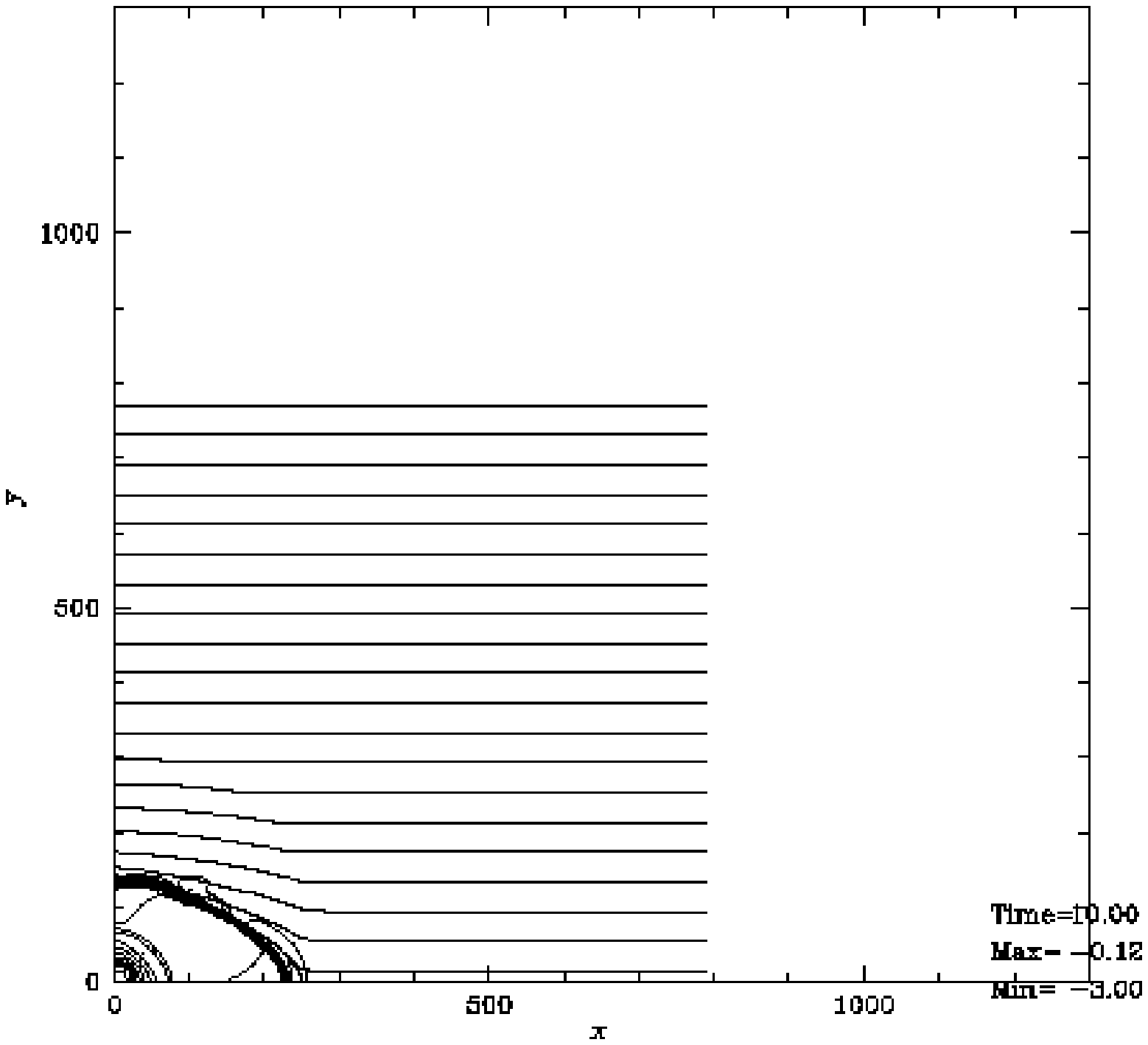}{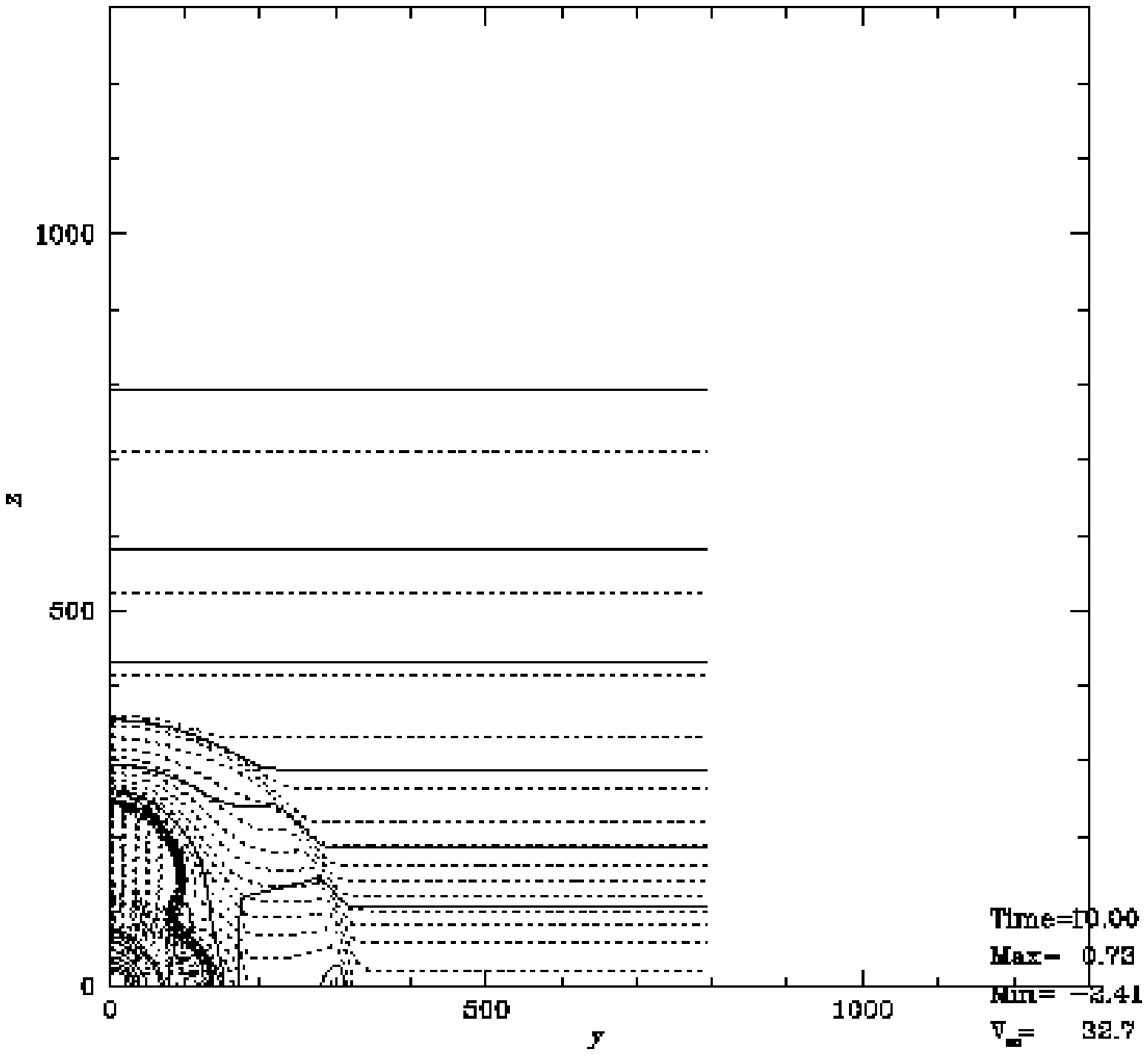}{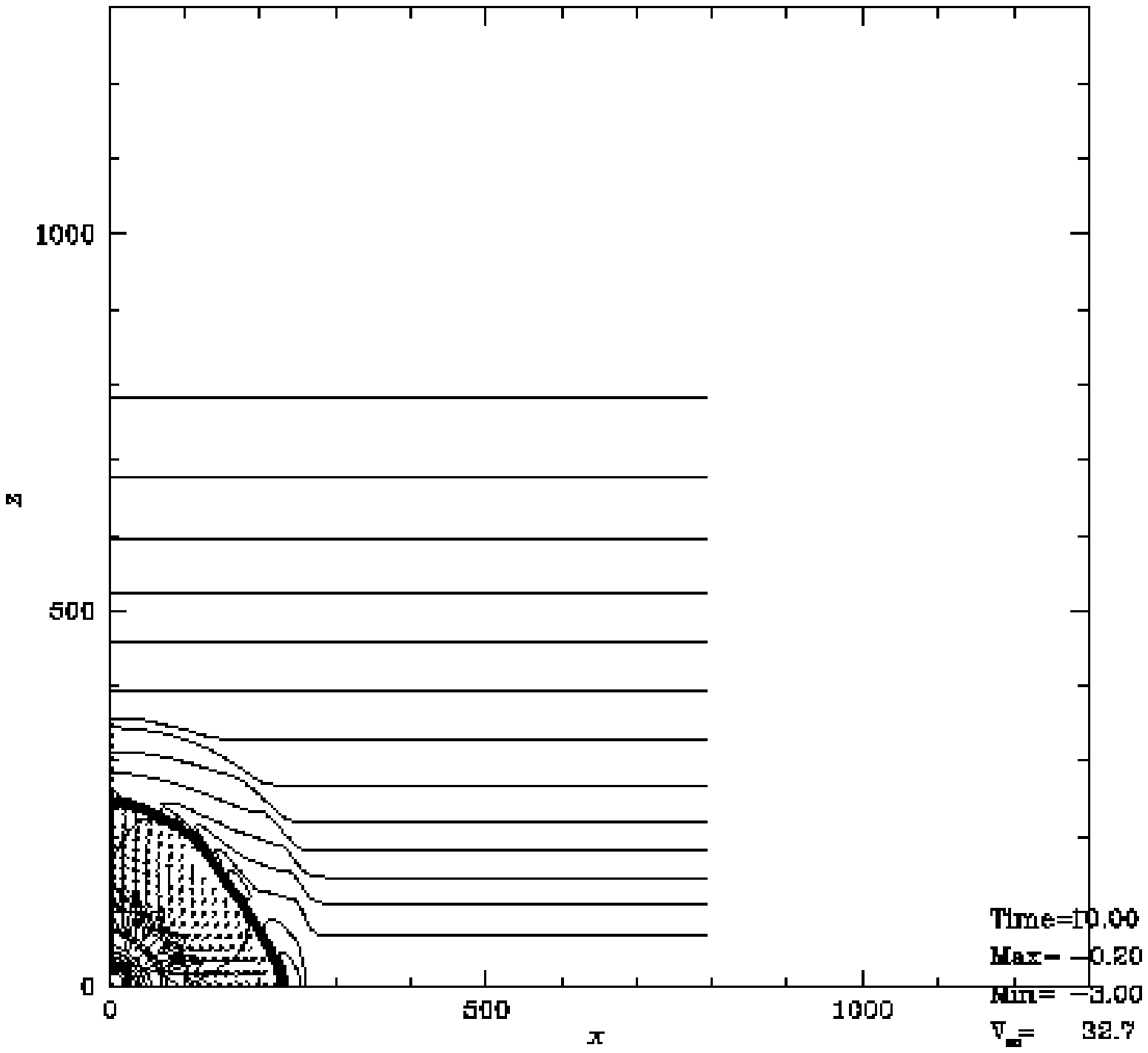}
\caption{The same as Fig.1 but for the model of $B_0\propto \rho^{1/2}$. 
Panels (a), (b) and (c) are the snapshots of $t=10$ Myr.
Panels (d), (e) and (f)  are those of $t=22.8$ Myr and 
 (g), (h) and (i) are for $t=30$ Myr.}
\end{figure*}
\clearpage

\setcounter{figure}{6}
\begin{figure*}
\plotthree{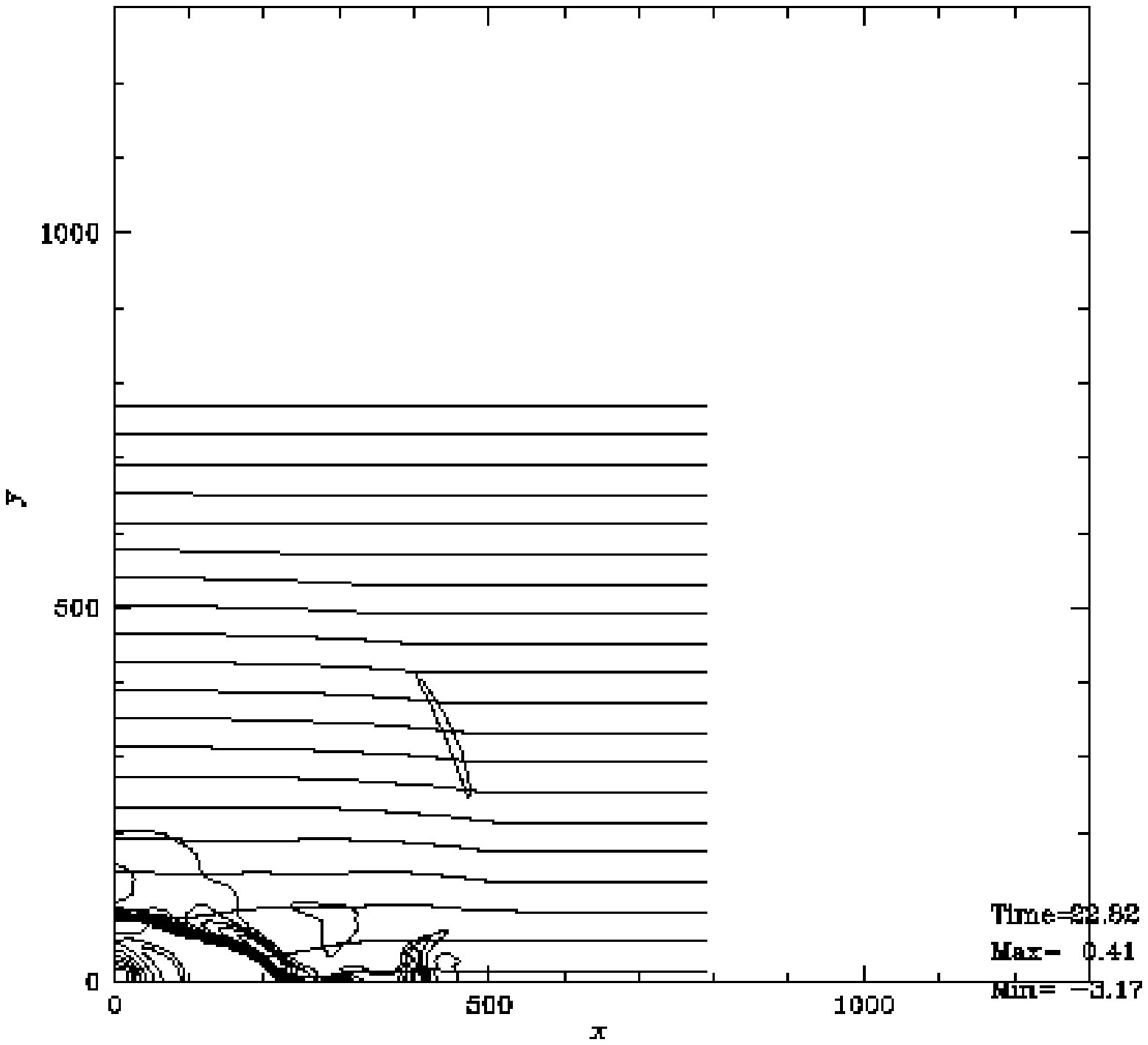}{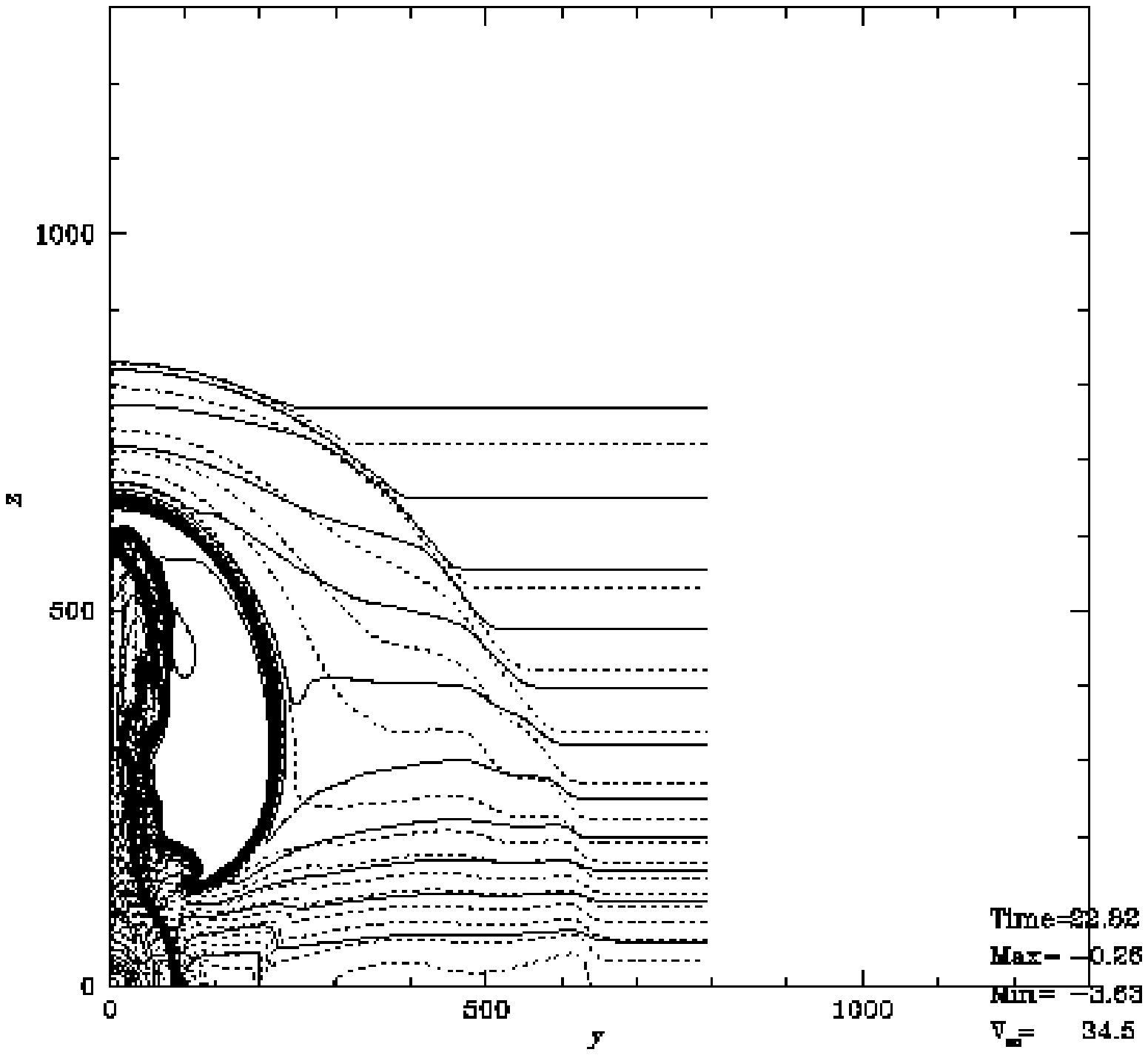}{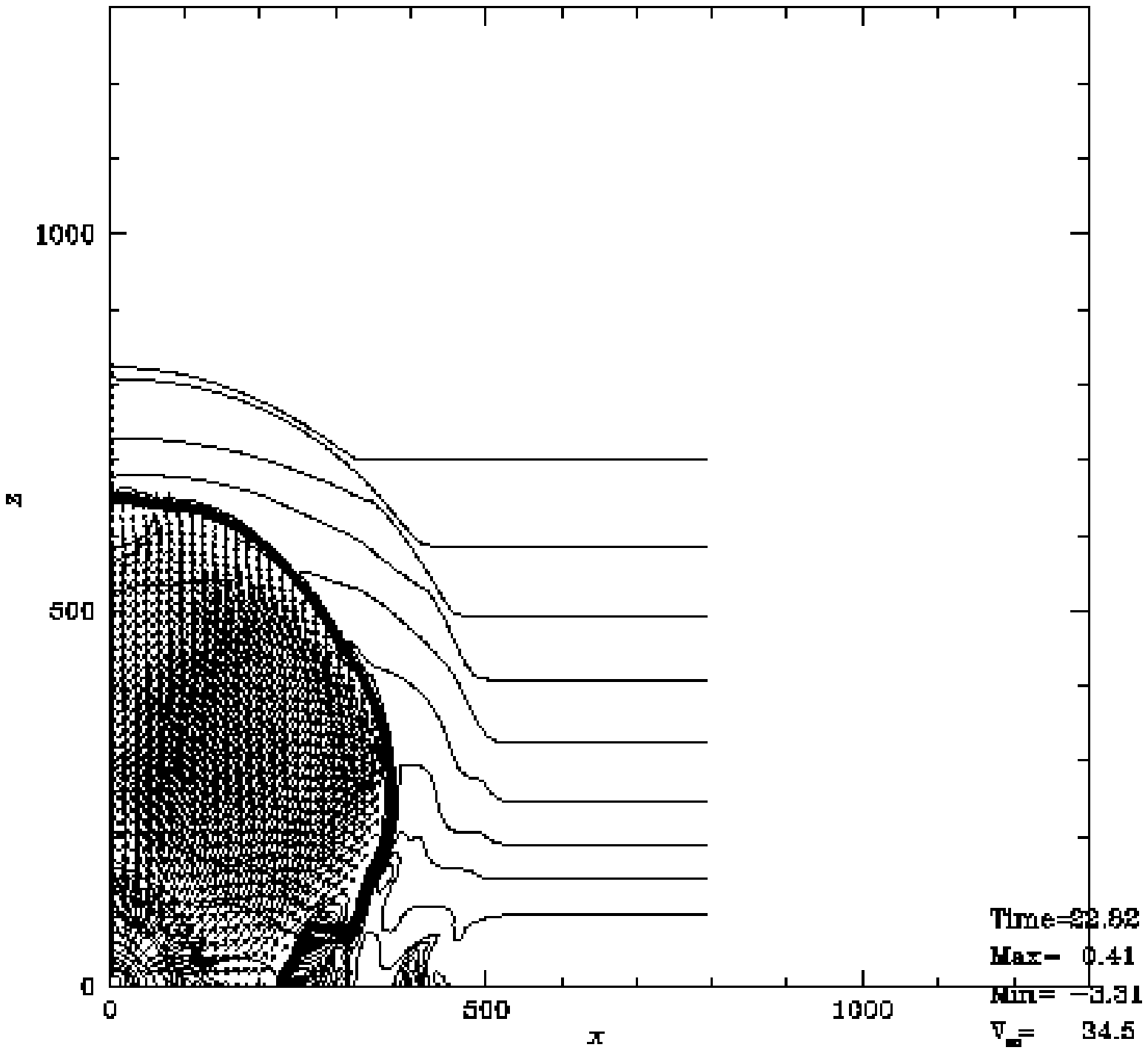}
\caption{continued.}
\end{figure*}
\clearpage

\setcounter{figure}{6}
\begin{figure*}
\plotthree{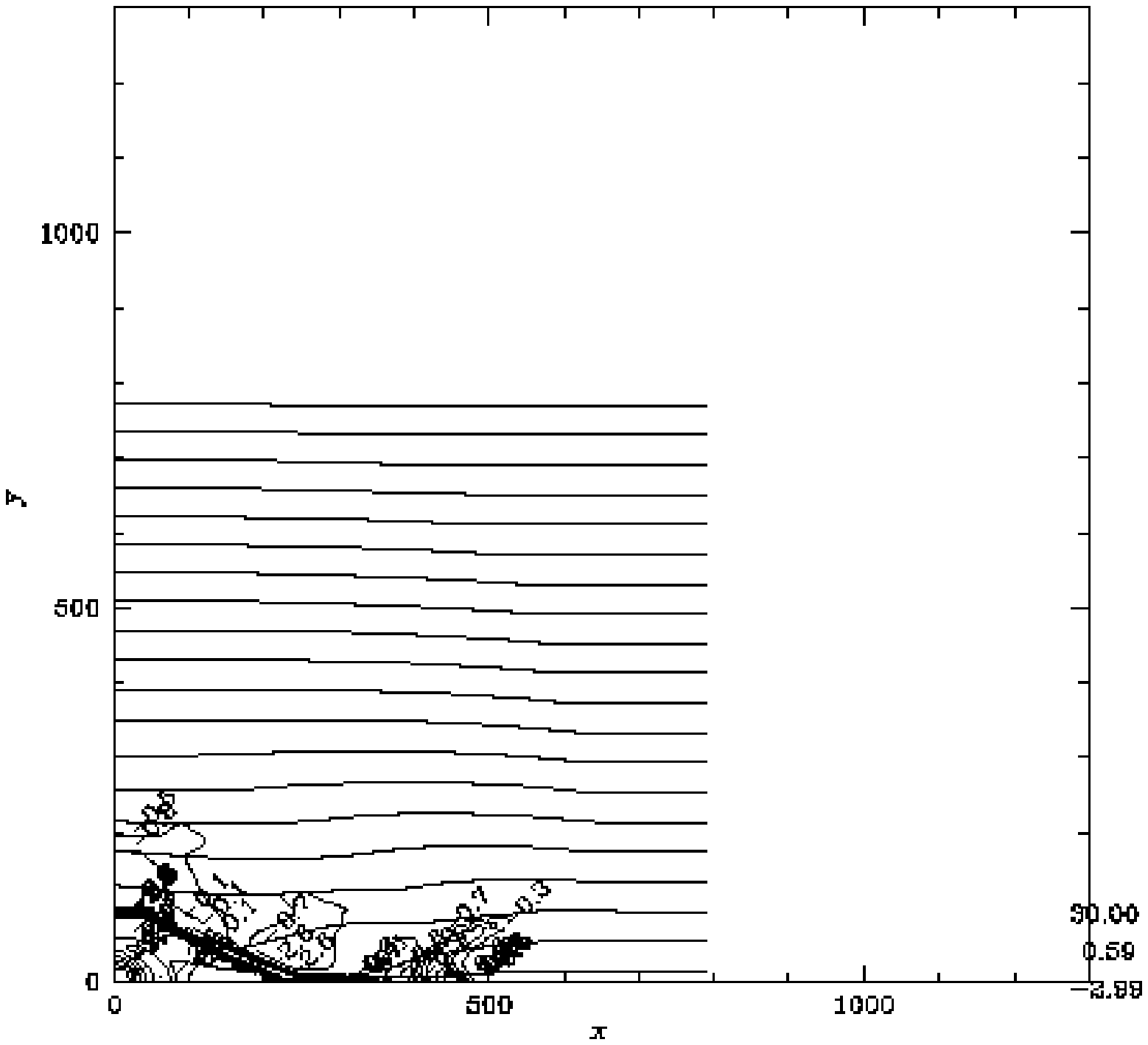}{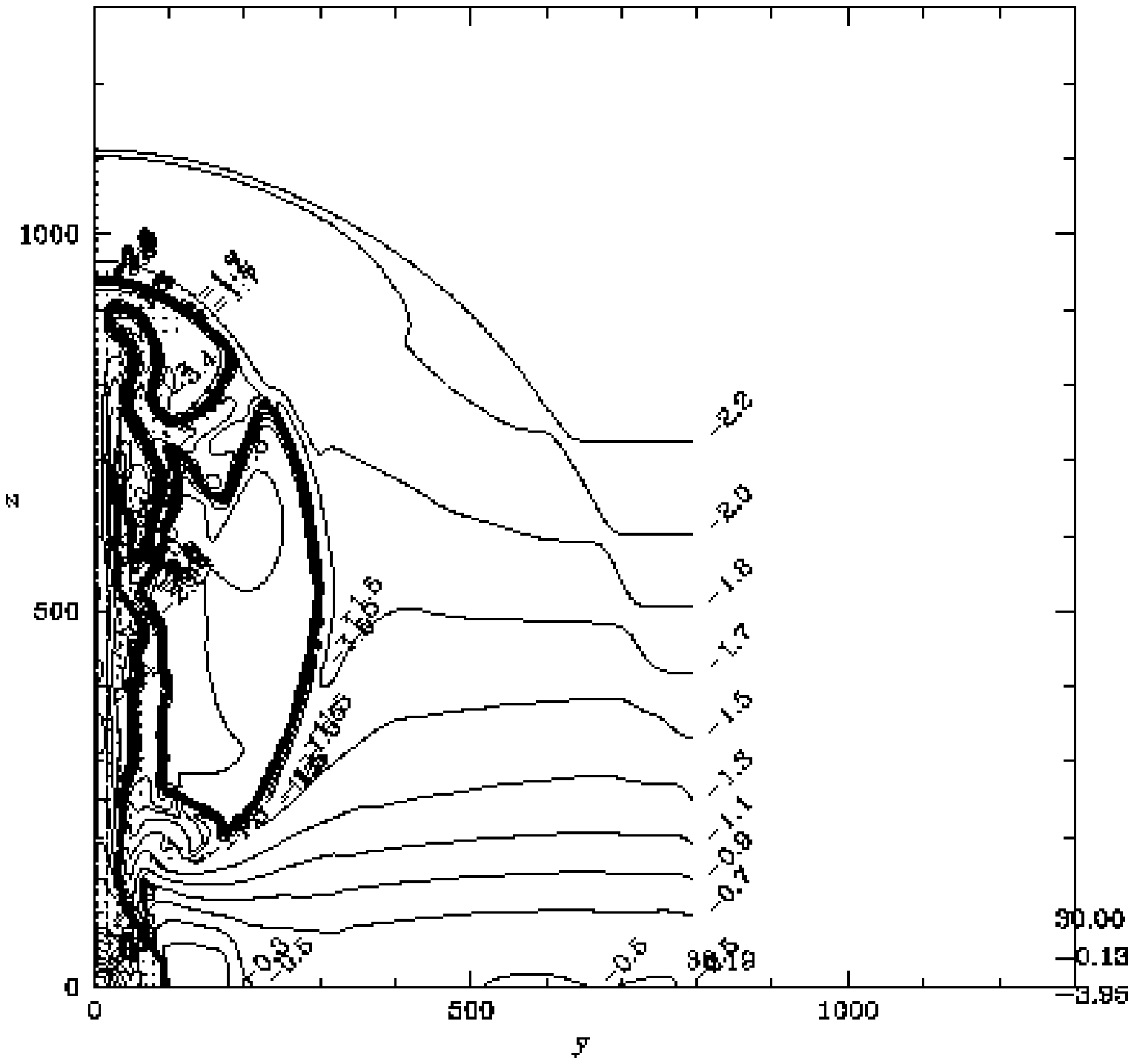}{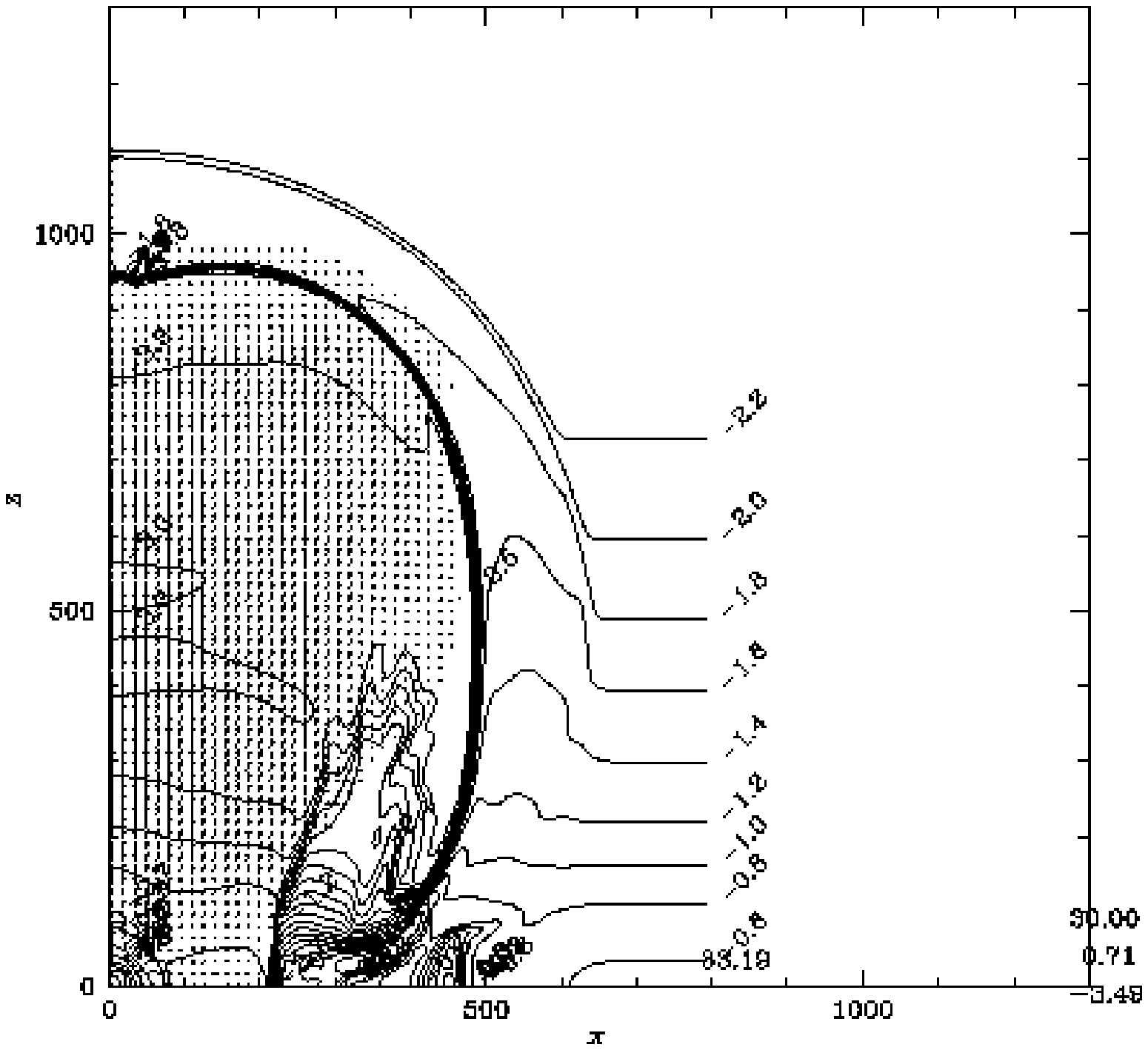}
\caption{continued.}
\end{figure*}

\setcounter{figure}{7}
\begin{figure*}
\plotthree{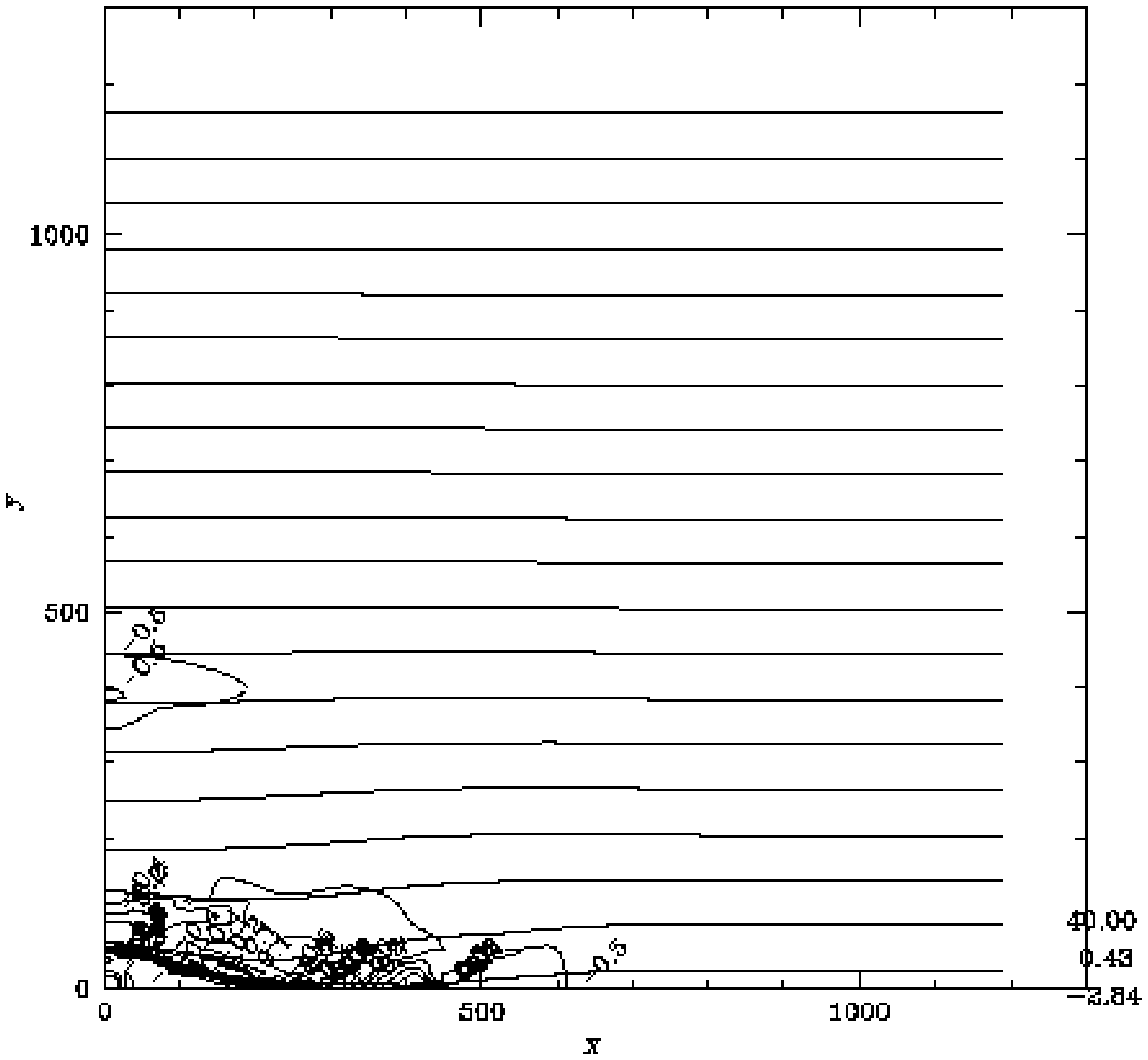}{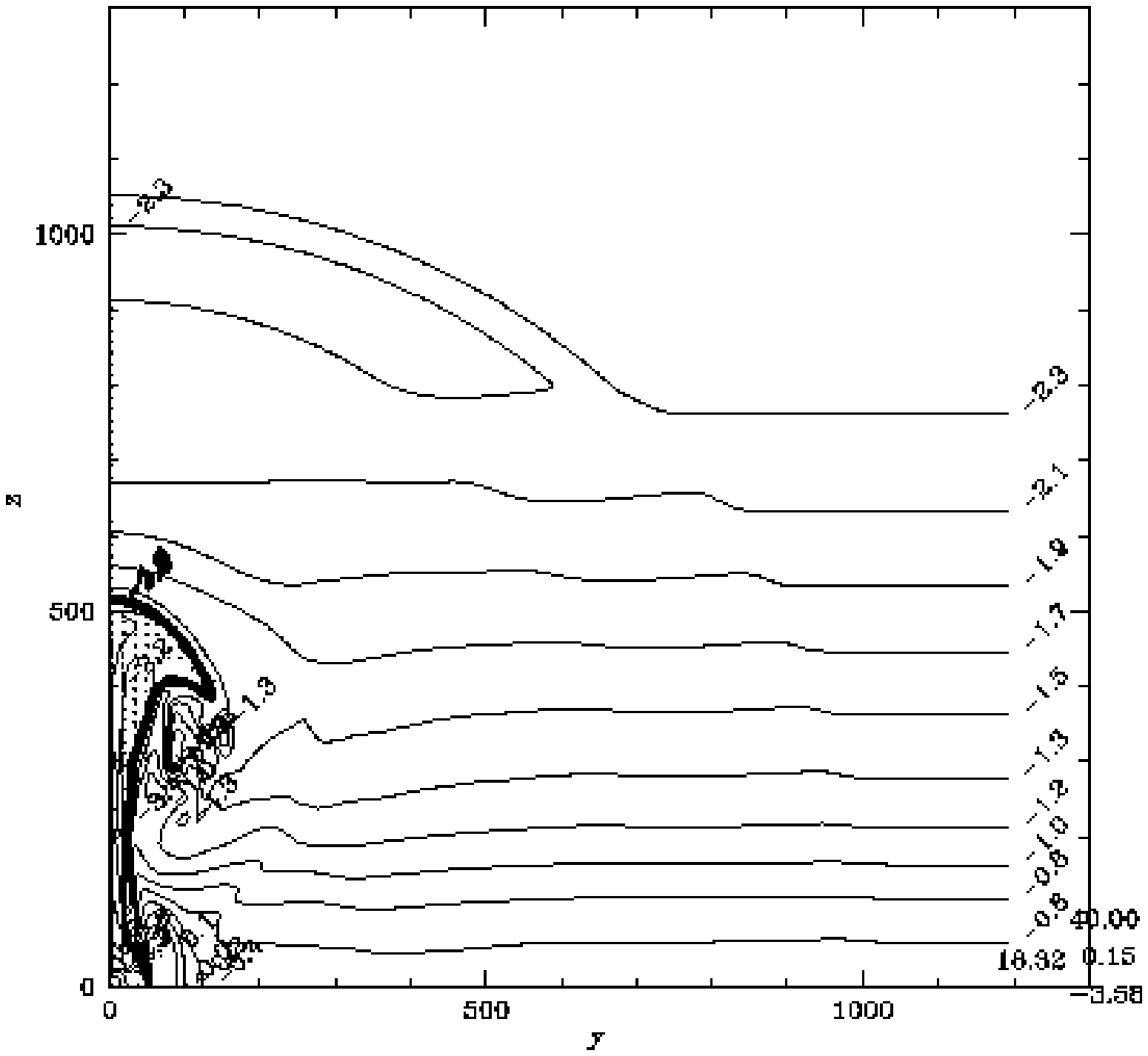}{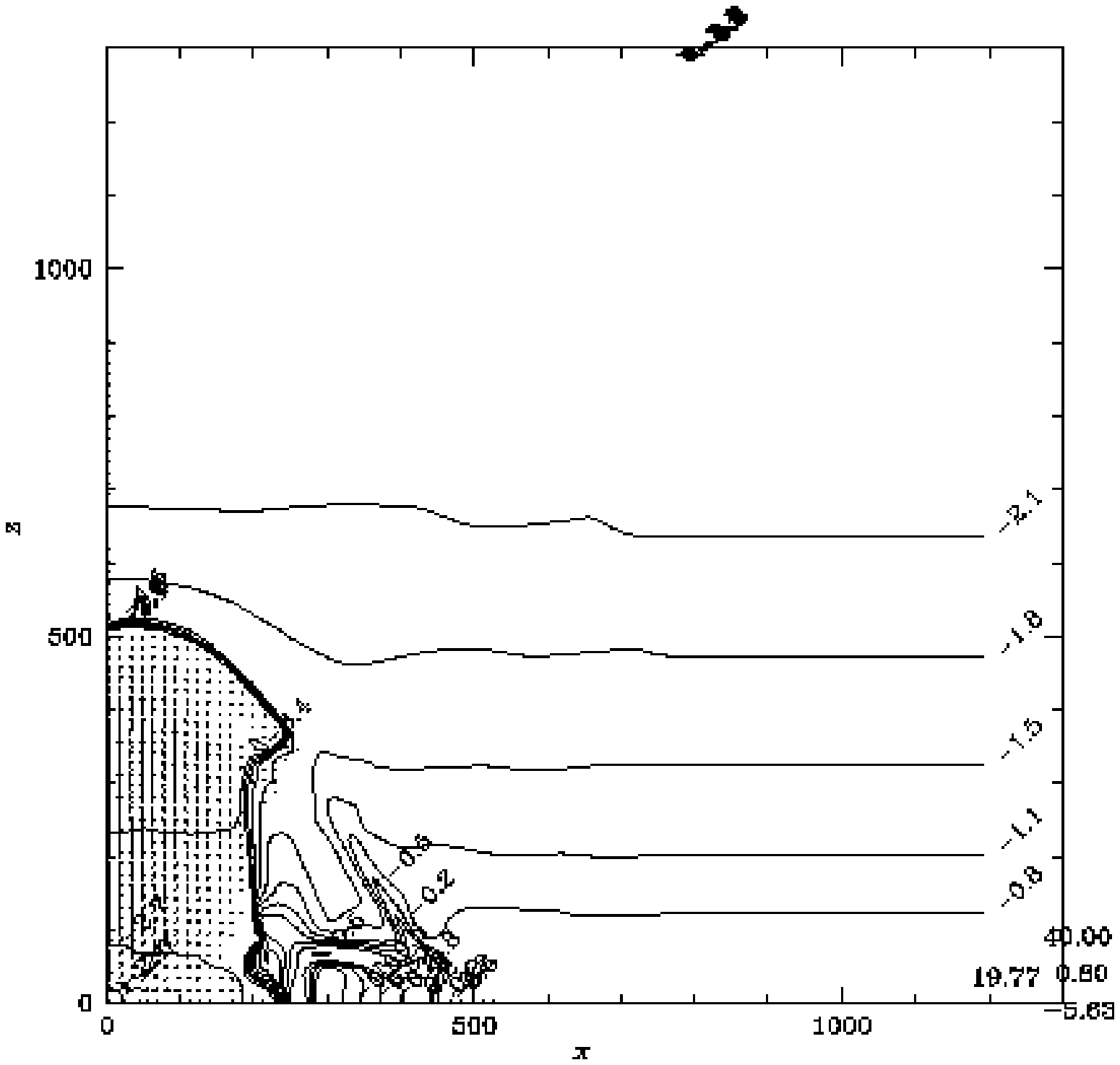}
\caption{The same as Fig.7 but for a model with a less mechanical luminosity
 of $L_{\rm SN}=10^{37}{\rm erg/s}$. 
The snapshot corresponds to $t=40$ Myr.}
\end{figure*}

\clearpage
\setcounter{figure}{8}
\begin{figure*}
\plotone{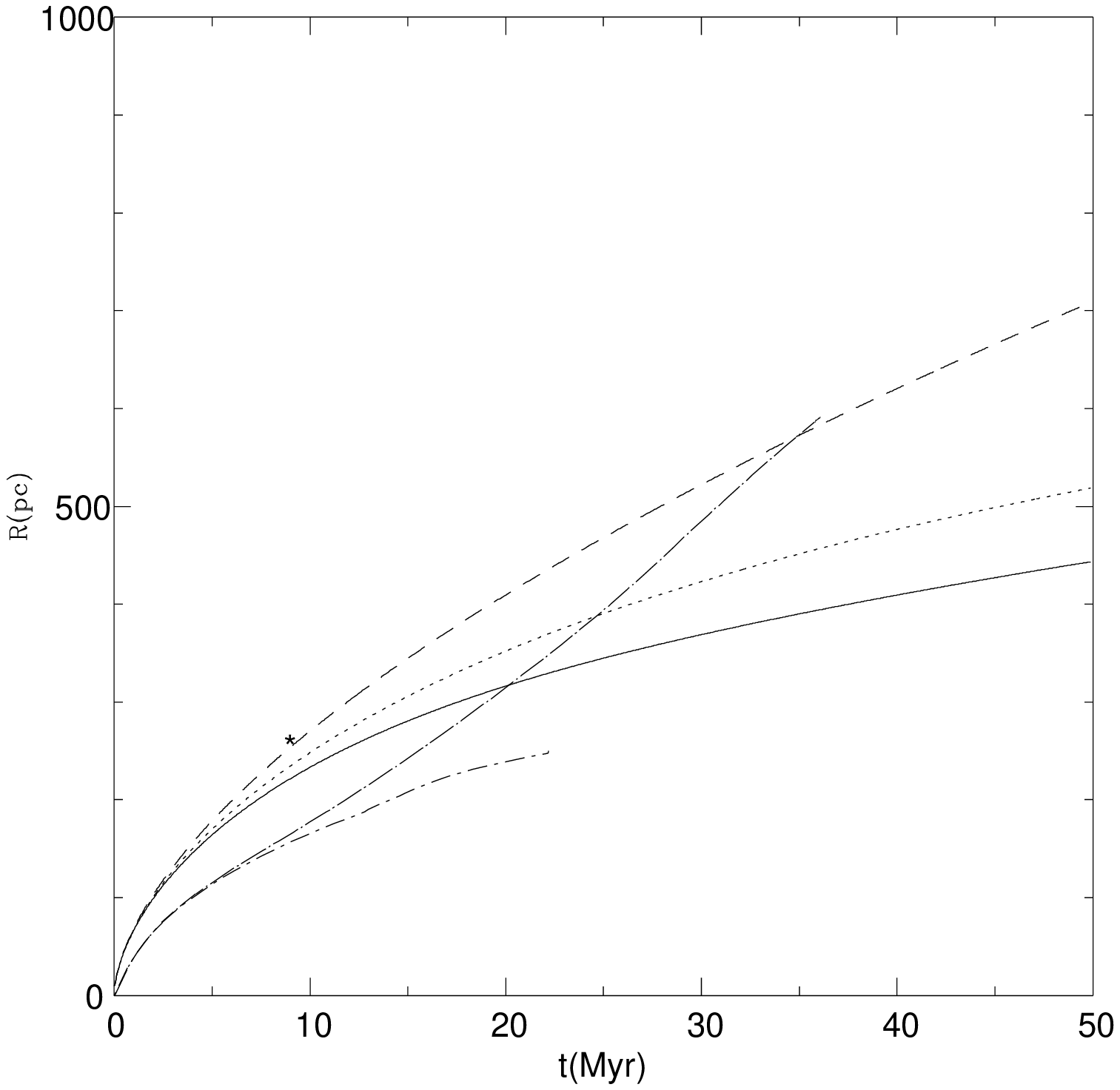}
\caption{Expansion law for a thin-shell model.
External interstellar pressures are chosen  equal to
$p_{\rm out}=1.7\times 10^{-12}{\rm erg~cm^{-3}}$ (solid line),
$p_{\rm out}=1\times 10^{-12}{\rm erg~cm^{-3}}$ (dotted line),
and $p_{\rm out}=0$ (dashed line).
The Weaver et al.'s (1977) solution is similar to a dashed curve and
 the point when the ram pressure is equal to the interstellar
 pressure with $p_{\rm out}=1.7\times 10^{-12}{\rm erg~cm^{-3}}$
 is also plotted by a asterisk.
The equivalent radii which are defined as 
 $r_{\rm equiv}=(3V_{\rm hot}/4\pi)^{1/3}$ for
 Models A and C are also illustrated with dash-dotted lines (lower:A;
 upper:C).}

\end{figure*}